\RequirePackage{fix-cm}
\documentclass[smallextended]{svjour3}       
\smartqed  
\usepackage{graphicx}
\usepackage{rotating}
 \usepackage{mathptmx}      
\usepackage{latexsym}
\usepackage{color, colortbl}

\usepackage{multirow}
\usepackage{hhline}

\usepackage{pifont}
\usepackage{mdwlist}
\usepackage{pdflscape}

\usepackage[usenames,dvipsnames]{xcolor}
\usepackage{cite}
\usepackage[authoryear]{natbib}

\usepackage{amsfonts}
\usepackage{amssymb}

\usepackage{wrapfig}
\usepackage{hyperref}
\usepackage{booktabs}

\definecolor{Gray}{gray}{0.7}

\newcommand{\cmark}{\ding{51}}%
\newcommand{\xmark}{\ding{55}}%

\begin{document}


\newcommand{\theHalgorithm}{\arabic{algorithm}}
\newcommand{\dfn}[1]{{\underline{#1}}}
\newcommand{\ksa}{K\_softAND}
\newcommand{\myK}{k}
\newcommand{\beq}{\begin{equation}}
\newcommand{\eeq}{\end{equation}}
\newcommand{\bit}{\begin{itemize}}
\newcommand{\eit}{\end{itemize}}
\newcommand{\goal}[1]{ {\noindent {$\Rightarrow$} \em {#1} } }
\newcommand{\hide}[1]{}
\newcommand{\comment}[1]{ {\footnotesize {#1} } }
\newtheorem{defn}{Definition}
\newcommand{\degp}{\emph{degree-preservation}~}
\newcommand{\degps}{\emph{degree-preservation}~}
\newcommand{\seedfinder}{\textbf{Seed-Finder}~}
\newcommand{\neighborexp}{\textbf{Neighbor-Expander}~}
\newcommand{\bridge}{\textbf{Bridge}~}
\newcommand{\FSGM}{\emph{G-Ray}}
\newcommand{\ESCD}{{\emph{Cor-Finder}}}
\newcommand{\weightedsoftand}{\textbf{WeightedSoftAND}~}
\newcommand{\Gq}{{\cal{H}}_q}
\newcommand{\Gt}{{\cal{H}}_t}
\newcommand{\rij}{{r_{i,j}}} 
\newcommand{\prop}{{\em{CePS}}}       
\newcommand{\propfull}{{{center-piece}}}       
\newcommand{\pmetric}{\emph{meeting probability}}
\newcommand{\ssp}{\emph{steady-state probability}}
\newcommand{\extract}{\emph{EXTRACT}}
\newcommand{\lqu}{``}
\newcommand{\rqu}{''}
\newcommand{\qi}{q_i}  
\newcommand{\pd}{pd}  
\newcommand{\QN}{{Q}}       
\newcommand{\Qset}{{\cal Q}}    
\newcommand{\Qsetm}{{\cal \acute{Q}}}    
\newcommand{\Qsetzero}{{\cal \varnothing}}    
\newcommand{\collision}{r}  
\newcommand{\cij}{ \collision_{i,j}}    
\newcommand{\cijpar}{ \collision{(i,j)}} 
\newcommand{\cqj}{ \collision ( \Qset, j)} 
\newcommand{\cqjK}{ \collision ( \Qset, j, \myK)} 
\newcommand{\cqjQ}{ \collision ( \Qset, j, \QN)} 
\newcommand{\cqjone}{ \collision ( \Qset, j, 1)} 
\newcommand{\cqHK}{ \collision ( \Qset, \Hgraph, \myK)} 
\newcommand{\Hgraph}{ { \cal H}} 
\newcommand{\mat}[1]{{\bf #1}}   
\newcommand{\matC}{ \mat{R}} 
\newcommand{\matW}{ \mat{W}} 
\newcommand{\matP}{ \mat{P}} 
\newcommand{\matS}{ \mat{S}} 
\newcommand{\matD}{ \mat{D}} 
\newcommand{\matQ}{ \mat{Q}} 
\newcommand{\matA}{\mat{A}}
\newcommand{\matX}{\mat{X}}
\newcommand{\matY}{\mat{Y}}
\newcommand{\matM}{ \mat{M}} 
\newcommand{\matdM}{\Delta\mat{M}}
\newcommand{\matI}{\mat{I}}
\newcommand{\matG}{\Ggraph}
\newcommand{\Ggraph}{ { \cal G} }
\newcommand{\matWnorm}{ \tilde{\mat{W}}} 
\newcommand{\matWnorma}{\tilde{{\mat{W}}}_{1}} 
\newcommand{\matWnormai}{\tilde{{\mat{W}}}_{1,i}} 
\newcommand{\matWnormb}{\tilde{{\mat{W}}}_{2}} 
\newcommand{\myB}{\mat{U}} 
\newcommand{\myC}{\mat{S}} 
\newcommand{\myD}{\mat{V}} 
\newcommand{\matlam}{\mat{{\Lambda}}}
\newcommand{\vece}{\vec{e}}
\newcommand{\prox}{{\rm{Prox}}}
\newcommand{\BitProxStatic}{Static-Bit-Proximity}
\newcommand{\bridgefind}{{{\em {NetShield}}}}
\newcommand{\ptrack}{\textbf{{pTrack}}}
\newcommand{\ctrack}{\textbf{{cTrack}}}
\newcommand{\FastOneUpdate}{\emph {{Fast-Single-Update}}}
\newcommand{\FastBatUpdate}{\emph {{Fast-Batch-Update}}}
\newcommand{\getqij}{\textbf{GetQij}}
\newcommand{\trackcentrality}{\emph {Track-Centrality}}
\newcommand{\trackprox}{\emph {Track-Proximity}}
\newcommand{\tracktopk}{\textbf{Track Top-K Queries}}
\newcommand{\FastRWR}{B\_LIN} 
\newcommand{\SimFastRWR}{NB\_LIN} 
\newcommand{\FastBitRWR}{BB\_LIN} 
\newcommand{\ItePow}{{\em OnTheFly}} 
\newcommand{\InvMat}{{\em PreCompute}} 
\newcommand{\Blk}{\textit{Blk}} 
\newcommand{\nrwr}{ProSIN}
\newcommand{\fastnrwr}{Fast-\nrwr}

\newcommand{\CorelFive}{\textbf{CoIR}} 
\newcommand{\CorelMMG}{\textbf{CoMMG}} 
\newcommand{\DblpAp}{\em {AP}} 
\newcommand{\DblpAc}{{\em AC}} 
\newcommand{\DblpAA}{{\em AA}}
\newcommand{\DblpPC}{\em {PC}}
\newcommand{\dtkarate}{{\em Karate}}
\newcommand{\dtnips}{\em {NIPS}}
\newcommand{\dtdatamining}{\em {DM}}
\newcommand{\dtmachinelearning}{\em {ML}}
\newcommand{\dtdblpdac}{{AC}}
\newcommand{\dtdblpdacposter}{{ACPost}}
\newcommand{\dtnetflix}{{\em {NetFlix}}}

\newcommand{\edge}{ {(j,l)}}    
\newcommand{\cqedge}{ \collision ( \Qset, \edge)} 
\newcommand{\cqedgeK}{ \collision ( \Qset, \edge, \myK)} 
\newcommand{\ciedge}{ \collision ( i, \edge)} 

\newcommand{\unitvi}{ \vec{e}_i} 

\newcommand{\matmon}{{\em Colibri}}
\newcommand{\matmonD}{{\em Colibri-D}} 
\newcommand{\matmonS}{{\em Colibri-S}} 
\newcommand{\lincur}{{\em Colibri-S}}

\newcommand{\bridging}{`{\em Shield-value}'}
\newcommand{\delLam}{\Delta \lambda}
\newcommand{\delLamH}{\Delta \hat{\lambda}}
\newcommand{\nchoosek}[2]{\left(\begin{array}{c}#1\\#2\end{array}\right)}
\newcommand{\backbone}{`{\em Shield-value}'}
\newcommand{\lamone}{\lambda}
\newcommand{\robust}{`{\em Vulnerability}'}
\newcommand{\vulg}{\textrm{V}(\mat{G})}

\newcommand{\arxiv}{{arXiv}}
\newcommand{\patents}{{\tt Patents}}
\newcommand{\authorpaper}{{\tt AuthorPaper}}
\newcommand{\autonsys}{{\tt AS}}

\newcommand{\dbar}{\bar{d}}
\newcommand{\Alpha}{\alpha}
\newcommand{\Beta}{\beta}
\newcommand{\tilda}{\lower.7ex\hbox{\textasciitilde}}

\newcommand{\curtrack}{{\textbf Incremental-CUR}}

\title{Graph-based Anomaly Detection and Description: A Survey
}


\author{Leman Akoglu         \and
        Hanghang Tong \and
        Danai Koutra }


\institute{Leman Akoglu \at
             Department of Computer Science,
             Stony Brook University,
             Stony Brook, NY 11794.\\
              Tel.: +1-631-632-9801,
              Fax: +1-631-632-2303. \\
              \email{leman@cs.stonybrook.edu}           
           \and
           Hanghang Tong \at
            Department of Computer Science,
               City College, City University of New York,
               New York, NY 10031 USA.\\
           \email{tong@cs.ccny.cuny.edu}
            \and
                      Danai Koutra \at
                       Computer Science Department,
                          Carnegie Mellon University,
                          Pittsburgh, PA 15217 USA.\\
                      \email{danai@cs.cmu.edu}
}

\date{Received: date / Accepted: date}

\maketitle

\begin{abstract}
Detecting anomalies in data is a vital
task, with numerous high-impact applications in areas such as security, finance, health care, and law enforcement.
While numerous techniques have been developed in past years for spotting
outliers and anomalies in unstructured collections of multi-dimensional points,
with graph data becoming ubiquitous, techniques for structured {\em graph} data have been of focus recently.
As objects in graphs have long-range correlations, a suite of novel technology has been developed for anomaly detection in graph data.

This survey aims to provide a general, comprehensive, and structured overview of the state-of-the-art methods for
anomaly detection in data represented as graphs.
As a key contribution, 
we give a general framework for the algorithms categorized under various settings: unsupervised vs. (semi-)supervised approaches, for static vs. dynamic graphs, for attributed vs. plain graphs. 
We highlight the effectiveness, scalability, generality, and robustness aspects of the methods. 
What is more, we stress the importance of anomaly {\em attribution} and highlight the major techniques that
facilitate digging out the root cause, or the `why', of the detected anomalies for further analysis and sense-making.
Finally, we present several real-world applications of graph-based anomaly detection in diverse domains, including financial, auction, computer traffic, and social networks. We conclude our survey with a discussion on open theoretical and practical challenges in the field.



\keywords{anomaly detection \and graph mining \and network outlier detection, event detection, change detection, fraud detection, anomaly description, visual analytics}
\end{abstract}

\section{Introduction}
\label{intro}


When analyzing large and complex datasets, knowing what stands out in the data is often at least,
or even more important and interesting than learning about its general structure. The branch of data mining concerned with discovering rare occurrences in datasets is called {\em anomaly detection}.
This problem domain has numerous high-impact
applications in security, finance, health care, law enforcement, and many others.

Examples include
detecting network intrusion or network failure \citep{Ide04,SunXZF08,conf/kdd/DingKBKC12},
credit card fraud \citep{Bolton01unsupervisedprofiling},
calling card and telecommunications fraud  \citep{journals/ida/CortesPV02,675496}, 
auto insurance fraud \citep{journals/sigkdd/PhuaAL04},
health insurance claim errors \citep{conf/kdd/KumarGM10},
accounting inefficiencies \citep{McGlohonBASF09},
email and Web spam \citep{conf/sigir/CastilloDGMS07},
opinion deception and reviews spam \citep{conf/www/OttCH12},
auction fraud \citep{PanditCWF07},
tax evasion \citep{journals/eswa/WuOLCY12,conf/kdd/AbeMPRJTBACKDG10},
customer activity monitoring and user profiling \citep{conf/kdd/FawcettP99,conf/kdd/FawcettP96},
click fraud \citep{Jansen08,Kshetri10},
securities fraud \citep{NevilleSJKPG05},
malicious cargo shipments \citep{conf/kdd/DasS07,conf/icdm/EberleH07}
malware/spyware detection \citep{conf/kdd/MaSSV09,provos07hotbots,DBLP:conf/sp/InvernizziC12},
false advertising \citep{socialspamlee},
data-center monitoring \citep{li2011thermocast},
insider threat \citep{conf/csiirw/EberleH09},
image/video surveillance \citep{conf/wiamis/DamnjanovicAI008,KrauszHerpers10mtaa},
 and many others.

 In addition to revealing suspicious behavior, anomaly detection is vital for spotting rare events, such as rare disease outbreaks 
or side effects in medical domain with vital applications in the medical diagnosis.
As ``one person's signal is another person's
noise'', yet another application of abnormality detection is data cleaning -- i.e., the removal of erroneous values or noise from data 
 as a pre-processing step
to learning more accurate models of the data.

\subsection{Outliers vs. Graph Anomalies}

\begin{figure}[!h]
\begin{center}
    \begin{tabular}{cc}
    \includegraphics[width=0.45\textwidth]{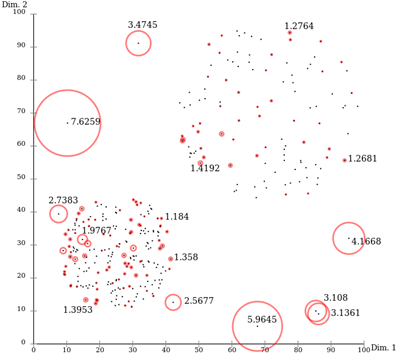}     &
    \includegraphics[width=0.45\textwidth]{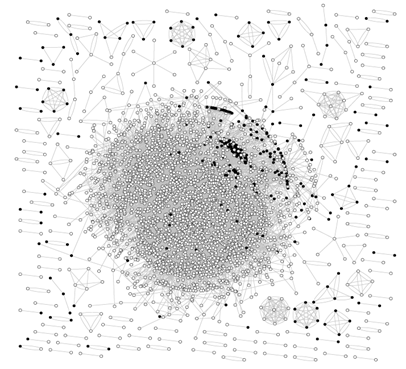}     \\\\
    (a) Clouds of points (multi-dimensional) &
    (b) Inter-linked objects (network) \\
    \end{tabular}
\caption{(a) Point-based outlier detection vs. (b) Graph-based anomaly detection.
}
\label{fig:compare}
\end{center}
\end{figure}

To tackle the abnormality detection problem, many techniques have been developed in the past decades, especially for spotting outliers and anomalies in unstructured collections of multi-dimensional data points.
On the other hand, data objects cannot always be treated as points lying in a multi-dimensional space independently.
In contrast, they may exhibit {\em inter-dependencies} which should be accounted for during the anomaly detection process (see Figure~\ref{fig:compare}).
In fact, data instances in a wide range of disciplines, such as physics, biology, social sciences, and information systems, are inherently related to one another.
Graphs provide a powerful machinery for
effectively capturing these long-range correlations among inter-dependent data objects.

To give an illustrative example, in a reviewer-product review graph data, the extent a reviewer is fraudulent depends on what ratings s/he gave to which products, as
well as how other reviewers rated the same products, to an extent how trustworthy their ratings are, which in turn again depends on what other products they rated, and so on. As can be seen, due to this long-range correlations in real-world datasets, detecting abnormalities in graph data is a significantly different task than that of detecting outlying points lying in a multi-dimensional feature space.
As a result, researchers have recently intensified their study of methods  for anomaly detection in  structured {\em graph} data.

\vspace{0.075in}
\noindent
\textbf{Why Graphs?}
\label{sec:why}
We highlight four main reasons that make graph-based approaches to anomaly detection vital and necessary: 

\begin{itemize}
\item {\em Inter-dependent nature of the data:}
As we briefly mentioned above, data objects are often related to each other and exhibit dependencies.
In fact, most relational data can be thought of as inter-dependent, which necessitates to account for related objects in finding anomalies.
Moreover, this type of datasets are abundant, including biological data such as the food web and protein-protein interaction (PPI) networks,
terrorist networks, email and phone-call networks, blog networks, retail networks, social networks,
to name but a few.

\item {\em Powerful representation:}
Graphs naturally represent the inter-dependencies by the introduction of links (or edges) between the related objects.
The multiple paths lying between these related objects effectively capture their long-range correlations.
Moreover, a graph representation facilitates the representation of rich datasets enabling the incorporation of node and edge attributes/types.

\item {\em Relational nature of problem domains:}
The nature of anomalies could exhibit themselves as relational.
For example in the fraud domain, one could imagine two types of scenarios: (1) {\em opportunistic} fraud that spreads by word-of-mouth (if one commits fraud, it is likely that his/her acquaintances will also do so), and (2) {\em organized} fraud that takes place by the close collaboration of a related group of subjects. Both of these scenarios point to relational treatment of anomalies.
Another example can be given in the performance monitoring domain, where the failure of a machine could cause the malfunction of
the machines dependent on it. Similarly,
the failure of a machine could be a good indicator of the possible other failures of machines in close spatial proximity to it (e.g., due to excessive increase of humidity in that particular region of a warehouse).

\item {\em Robust machinery:} Finally, one could argue that graphs serve as more adversarially-robust tools. For example in fraud detection systems, behavioral clues such as log-in times and locations (e.g. IP addresses) can be easily altered or faked by advanced fraudsters. On the other hand, it may be reasonable to argue that the fraudsters could not have a global view of the entire network (e.g. money transfer, telecommunication, email, review network) that they are operating in. As such, it would be harder for a fraudster to {\em fit in} to this network as good as possible without knowing its entire characteristic structure and dynamic operations.

\end{itemize}

\subsection{Challenges}

We first discuss the very immediate challenge associated with our problem of interest. It stems from the fact that no unique definition for the problem of {\em anomaly detection} exists. 
The reason is that the general definition of an anomaly or an outlier is a vague one: the definition becomes meaningful only under a given context or application.
The very first definition of an outlier dates back to 1980, and is given by Douglas M. Hawkins \citep{hawkins:1980}:

\begin{definition}[Hawkins' Definition of Outlier, 1980]
``An outlier is an observation that differs  so much from other observations as to 	arouse suspicion that it was generated by a different mechanism.''
\end{definition}

As one notices, the above definition is quite general and thus make the detection problem an open-ended one. As a result, the problem of anomaly detection has been defined in various ways in different contexts. In other words, the problem has many definitions often tailored for the specific application domain, and also exhibits various names such as outlier, anomaly, outbreak, event, change, fraud, detection, etc. In some applications, such as data cleaning, outliers are even called the noise---``one man's signal is another man's noise''.
Nevertheless, anomaly detection is one of the most evident problems in data mining with numerous applications, and the field of anomaly detection itself is well established.

Following the general definition of an outlier by Hawkins as given above, we provide a general definition for the graph anomaly detection problem as follows.

\begin{definition}[General Graph Anomaly Detection Problem] \\
\indent \textbf{Given} a (plain/attributed, static/dynamic) graph database, \\
\indent \textbf{Find} the graph objects (nodes/edges/substructures) that are rare and 
 that differ 
 \indent \indent significantly from the majority of the reference objects in the graph.
\end{definition}

For practical purposes, a record/point/graph-object is flagged as anomalous 
if its rarity/likelihood/outlierness score exceeds a user-defined or an estimated threshold.
In other words, an anomaly is treated as a data object or a group of objects that is
rare (e.g., rare combination of categorical attribute values),
isolated (e.g., far-away points in $n$-dimensional spaces), and/or
surprising (e.g., data instances that do not fit well in our mental/statistical model, or need too many bits to describe under the Minimum Description Length principle \citep{rissanen99}).

Next, we discuss the challenges associated with anomaly detection and attribution, which can be grouped into two: (1) data-specific, and (2) problem-specific challenges. We also specifically highlight the challenges associated with graph-based anomaly detection.


\vspace{0.075in}
\noindent
\textbf{Data-specific challenges:} Simply put, the challenges with respect to data are those of working with big data; namely volume, velocity, and variety of massive, streaming, and complex datasets.
The same challenges generalize to graph data as well. 
\vspace{0.05in}

 {\em Scale and Dynamics}:  With the advance of technology, it is much easier than was in the past to collect and analyze very large datasets.
As of today, Facebook (graph) consists of more than a billion users\footnote{\url{http://newsroom.fb.com/Key-Facts}} (i.e., nodes), the Web (graph) contains more than 40 billion pages\footnote{\url{http://www.worldwidewebsize.com/}},
and over 6 billion users own a cell phone\footnote{\url{http://huff.to/Rc2vbU}} which makes the telecommunication networks billion-scale graphs.
Not only is the size of real data in tera- to peta-bytes, but also the rate at which it arrives is high.
Facebook users generate billions of objects (e.g. posts, image/video uploads, etc.),
billions of credit card transactions are performed every day, billions of click-through traces of Web users are generated each day, and so on. This kind of data generation can be thought of as streaming graph data.

 {\em Complexity}: In addition to (graph) data size and dynamicity, the datasets are rich and complex in content; including for example user demographics, interests, roles, as well as different types of relations.
As such, incorporation of these additional information sources makes the graph representation a complex one, where nodes and edges can be typed, and have a long list of attributes associated with them.

As a result, methods  which could scale to very large graphs, update their estimations when the graph changes over time, and that could effectively incorporate all the available and useful data sources are essential for graph-based anomaly detection.

\vspace{0.075in}
\noindent
{\bf Problem-specific challenges:} Additional challenges arise with respect to the anomaly detection task itself. 
\vspace{0.05in}

\noindent

 {\em Lack and Noise of Labels}: One main challenge is that the data often comes without any class labels, that is, the ground truth of which data instances are anomalous and non-anomalous does not exist. Importantly, the task of manual labeling is quite challenging given the size of the data. To make things worse, even though endless human power were available, due to the complexity of certain labeling tasks, the labels are expected to be noisy and of varying quality depending on the annotator. According to Nobel laureate Daniel Kahneman ``humans are incorrigibly inconsistent in making summary judgments of complex information'' \cite{kahneman2011thinking}. Surprisingly, they frequently give different answers when asked to evaluate the same information twice. For example, experienced radiologists who evaluate chest X-rays as normal and abnormal are found to contradict themselves  20\% of the time when they see the same picture on separate occasions.

Due to challenges in obtaining labels, supervised machine learning algorithms are less attractive for the task of anomaly detection.
It has been shown that humans can perform at best as good as random in labeling a review as fake or not, just by looking at its text \citep{Ott12} but can potentially do better by analyzing other relevant information such as the authors of the review. Likewise, a single transaction could be treated as anomalous only in relation to a history of previous transactions.  
These indicate that additional resources and information are needed to obtain human labels, which makes it costly to acquire them and harder and more time-consuming for the human annotators to sort through.
What is more, the lack of true labels, i.e. ground truth data, also makes the evaluation of anomaly detection techniques challenging.

 {\em Class Imbalance and Asymmetric Error}: The second challenge arises due to the unbalanced nature of the data; since anomalies are rare only a very small fraction of  the data is expected to be abnormal. Moreover, the cost of mislabeling a good data instance versus a bad instance may change depending on the application, and further could be hard to estimate beforehand. For example, mislabeling a cancer patient as healthy could cause fatal consequences while mislabeling an honest customer as a fraudster could cause loss of customer fidelity.
If learning-based techniques are to be employed, those issues regarding class imbalance and asymmetric error costs should be carefully accounted for.

 {\em Novel Anomalies}:
The third point is the wrist-fight nature of the problem setting, especially in the fraud detection domain.
The more the fraudsters understand the ways the detection algorithms work, the more they change their techniques in a way to by-pass the detection and fit-in to the norm. As a result, not only the algorithms should be adaptive to changing and growing data over time, they should also be adaptive to and be able to detect novel anomalies in face of adversaries.

 {\em ``Explaining-away'' the Anomalies}:
Additional challenges lie in explaining the anomalies in the post-detection phase. This involves either digging out the root cause of an anomaly, telling a coherent story for the `why' and `how' of the anomaly, and/or presenting the results in a user-friendly form for further analysis. Most of the existing detection techniques, while doing a reasonably good job in spotting the anomalies, completely leave out this description or attribution phase and thus make it hard for humans to make sense of the outcome.

\vspace{0.075in}
\noindent
{\bf Graph-specific challenges:} All of the above challenges associated with the anomaly detection problem generalize to graph data. 
Graph-based anomaly detection, on the other hand, has additional challenges as well.
\vspace{0.05in}

 {\em Inter-dependent Objects}: Firstly, the relational nature of the data makes it challenging to quantify the anomalousness of graph objects. While in traditional outlier detection, the objects or data points are treated as independent and identically distributed (i.i.d.) from each other,  the objects in graph data have long-range correlations. Thus, the ``spreading activation'' of anomalousness or ``guilt by associations'' need to be carefully accounted for.

 {\em Variety of Definitions}:
Secondly, the definitions of anomalies in graphs are much more diverse than in traditional outlier detection, given the rich representation of graphs. For example, novel types of anomalies related to graph substructures are of interest for many applications, e.g., money-laundering rings in trading networks. 

 {\em Size of Search Space}:
The main challenge associated with more complex anomalies such as graph substructures is that the
search space is huge, as in many graph theoretical problems associated with graph search. The enumeration of possible substructures is combinatorial which makes the problem of finding out the anomalies a much harder task. This search space is enlarged even more when the graphs are attributed as the possibilities span both the graph structure and the attribute space. As a result, the graph-based anomaly detection algorithms need to be designed not only for effectiveness but also for efficiency and scalability.


\subsection{Previous Surveys and Our Contributions}
There exist very comprehensive
survey articles on anomaly and outlier detection in general that focus on points of multi-dimensional data instances.
In particular, \citep{Chandola2009} covers outlier detection techniques, \citep{ZimekSK12} focuses on outlier detection in high dimensions, and \citep{Schubert12} deals with local outlier detection techniques.
In addition, survey and special issue journal articles that address anomaly, event, and change detection include \citep{ChandolaBK12,MargineantuWD10,journals/tip/RadkeAAR05}.
Finally, due to the wide-range of application domains, fraud detection has attracted many surveys \citep{FlegelVB10,fraudcorr10,EdgeS09}.

 None of the previous surveys, however, discuss the anomaly detection problems in the particular context when one is confronted with large graph datasets. Further, they also do not focus, at least not directly, on graph-based detection techniques. Therefore,
 in this survey we aim to provide a comprehensive and structured overview of the state-of-the-art techniques for anomaly, event, and fraud
 detection in data represented as graphs.
As such, our focus is notably different from, while being complementary to the earlier surveys.
Specifically,  our contributions are listed as follows.
 \begin{enumerate}
 \item Different from previous surveys on anomaly and outlier detection, we focus on abnormality detection in (large)
 graph datasets, using graph-based techniques.
 \item We comprehensively explore unsupervised techniques that exploit the graph structure, as well as (semi-) supervised methods that employ relational learning. 
 \item We put the abnormality (anomaly, event, fraud) detection methods under
 a unifying lens, point out their connections, pros and cons (e.g., scalability, robustness, generality, etc.) and applications on diverse real-world tasks.
 \item In addition to anomaly detection, we highlight the importance of explaining the detected anomalies and provide a survey of analysis tools and techniques for post-detection exploration and sense-making.
 \end{enumerate}

\subsection{Overview and Organization} 

We present our survey in four major parts.  A general outline and a list of topics we cover
are given as follows. 

\begin{itemize}
\item[I.] \textbf{Anomaly detection in static graphs} (Section 2)
		\begin{itemize}

		\item[(a)]  Anomalies in plain (unlabeled) graphs
		\item[(b)]  Anomalies in attributed (node-/edge-labeled) graphs
		\end{itemize}
		
\item[II.] \textbf{Anomaly detection in dynamic graphs} (Section 3)
		\begin{itemize}
		
		\item[(a)] Feature-based events
		\item[(b)] Decomposition-based events
		\item[(c)] Community- or clustering-based events
		\item[(d)] Window-based events
		
%
		\end{itemize}
		
\item[III.]  \textbf{Graph-based anomaly description} (Section 4)
		\begin{itemize}
		
			\item[(a)]  Interpretation-friendly graph anomaly detection
			\item[(b)]  Interactive graph querying and sense making

		
%
		\end{itemize}		

\item[IV.]  \textbf{{Graph}-based anomaly detection in real-world applications} (Section 5)

\begin{tabular}{l  l}
\hspace{-0.15in} (a) Anomalies in telecom networks & \hspace{-0.1cm}     (f) Anomalies in opinion networks \\
\hspace{-0.15in} (b) Anomalies in auction networks & \hspace{-0.1cm}     (g) Anomalies in the Web network \\
\hspace{-0.15in} (c) Anomalies in account networks & \hspace{-0.1cm}     (h) Anomalies in social networks \\
\hspace{-0.15in} (d) Anomalies in security networks & \hspace{-0.1cm}     (i) Anomalies in computer networks \\
\hspace{-0.15in} (e) Anomalies in financial networks & \hspace{-0.1cm}     \\
\end{tabular}

\end{itemize}

The first part focuses on anomaly detection methods for
{\em static} graph data, and is covered for both unlabeled (plain) and labeled (attributed) graphs.
The second part focuses on change or event detection approaches for  time-varying or {\em dynamic} graph data, based on edit distances and connectivity structure. 
The overview of the first two sections, along with the areas with open problems and challenges are provided in Table~\ref{tab:summary23-overview}.
In the third part, we stress the importance of anomaly {\em attribution} in revealing the root-cause of the detected anomalies and in presenting anomalies in a user-friendly form. We provide the state-of-the-art tools that could facilitate the post-analysis of detected anomalies for the crucial task of sense-making.
Finally, in the fourth and last part we demonstrate graph-based anomaly detection techniques in action, where we discuss several real-world \textit{applications} in diverse domains.

\begin{table}[!h]
\vspace{-0.1in}
\centering{
\caption{\normalsize Categorization of graph-based techniques in Section \ref{sec:static} and Section \ref{sec:dynamic}.} 
\label{tab:summary23-overview}
{\small
\begin{tabular}{|l||c|c|} \hline
& {\normalsize \textbf{Plain}} & {\normalsize \textbf{Attributed}} \\ \hline \hline
\multirow{3}{*}{{\normalsize \textbf{Static}}} & & \\
&  {\small \textbf{[Section \ref{sec:staticplain}]}}  &  {\small \textbf{[Section \ref{sec:staticattributed}]}}  \\ \cline{2-3}
& \multicolumn{2}{c|}{\multirow{1}{*}{{\small{\textit{\textbf{Open}}}}} }\\
\hline \hline
\multirow{4}{*}{{\normalsize \textbf{Dynamic}}} & & \\
&  {\small \textbf{[Section \ref{sec:timeEvolvingGraphs}]}}  & {\small \textbf{[Section \ref{sec:timeEvolvingGraphs}]}} \\
&  & \small{\textbf{\textit{Many Open}}}\\
&  & \small{\textbf{\textit{Challenges}}}\\ \hline
\end{tabular}
}
}
\vspace{-0.1in}
\end{table}

\begin{figure}[!h]
\begin{center}
    \begin{tabular}{c}
    \includegraphics[width=\textwidth]{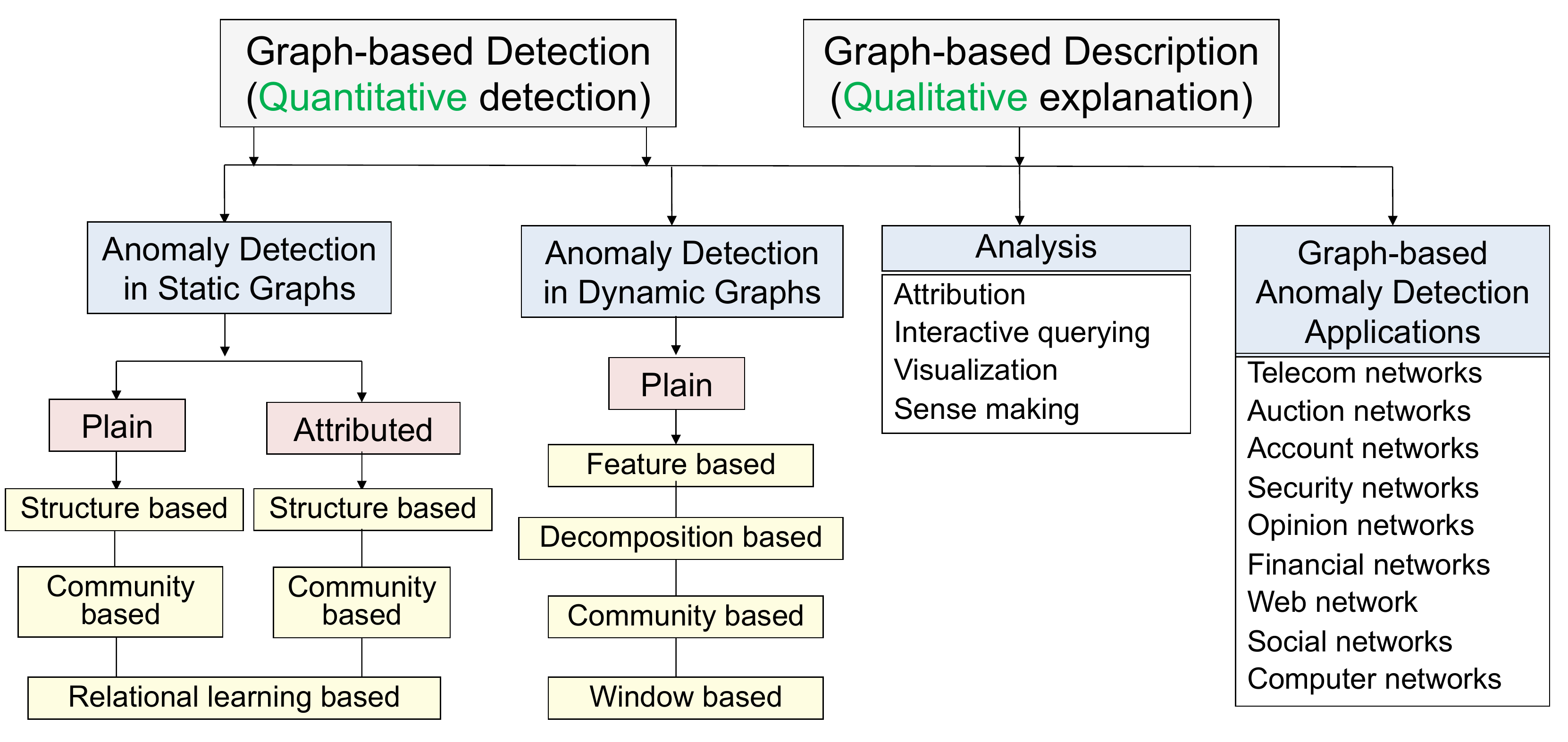}    \\\\
    \end{tabular}
\caption{Graph-anomaly detection: the outline of the survey.
}
\label{fig:survey}
\end{center}
	\vspace{-0.15in}
\end{figure}

We show the general outline of our survey in Figure~\ref{fig:survey} illustrating a sketch of the taxonomy.
In the first two parts, namely static and dynamic graph anomalies, we focus on {\em unsupervised} 
techniques as well as {\em (semi-) supervised} approaches based on relational classification.
Later in the third part, we focus on {\em qualitative analysis} techniques for the sense-making of spotted anomalies.
Finally in part four, we present a long list of {\em applications} of graph-based anomaly detection in a wide range of networks, including finance, security, accounting, to name a few.

\definecolor{Gray}{gray}{0.9}
\newcolumntype{g}{>{\columncolor{Gray}}c}

%


\section{Anomaly Detection in Static Graphs}
\label{sec:static}

In this section, we will address the anomaly detection in static snapshots of graphs. That is, the main task here is to spot anomalous network entities (e.g., nodes, edges, subgraphs) given the entire graph structure.
We start with a very brief overview of outlier detection techniques in static clouds of data points and provide pointers for further reading. Next, we survey anomaly detection techniques for static graphs.


\vspace{0.15in}
\noindent \textbf{Overview: Outliers in Clouds of Data Points}
\vspace{0.1in}

\sloppy{
Outlier detection deals with the problem of spotting outlying points in the (high-dimensional) feature space of data points.
While not directly related, outlier detection techniques are employed in graph-based anomaly detection, for example after a graph-feature extraction step as we describe in this section.
Thus it is beneficial to know of general outlier detection methods for spotting graph anomalies.

In outlier detection, some methods provide binary 0/1 classification of data points, i.e. outlier vs. non-outlier, while most methods try to assign what is called an outlierness score that enables the quantification of the level of outlierness of the objects and subsequently rank the objects accordingly. For an illustration, see Figure \ref{fig:compare}(a).
}

There are several different ways of multi-dimensional outlier detection. The techniques can be classified into
density-based \citep{conf/sigmod/BreunigKNS00,conf/icde/PapadimitriouKGF03},
distance-based \citep{conf/vldb/KnorrN98,DBLP:journals/pvldb/OrairTWMP10,DBLP:journals/datamine/GhotingPO08,DBLP:conf/icde/WangPT11,conf/sigmod/AggarwalY01,conf/dmkd/ChaudharySM02},
depth-based \citep{Ruts1996},
distribution-based \citep{journals/informaticaLT/Saltenis04},
clustering-based \citep{journals/prl/HeXD03,journals/pami/MillerB03,conf/cisis-spain/LietoOP08,DBLP:conf/incdm/WangWW12},
classification-based \citep{conf/kdd/AbeZL06,Hempstalk2008,conf/icmla/JanssensFP09},
information theory-based \citep{DBLP:conf/sdm/SmetsV11,COCO1557042,DBLP:conf/icdm/Ando07},
spectrum-based \citep{DBLP:journals/kais/LiuXCHD13}, and
subspace-based \citep{conf/icdm/MullerAIMB12,conf/cikm/MullerSS10,DBLP:conf/icdm/KriegelKSZ12,DBLP:conf/icde/KellerMB12} techniques.
Moreover, there exist outlier detection techniques that can work with categorical features \citep{conf/kdd/DasS07,DBLP:conf/sdm/SmetsV11,DBLP:conf/cikm/AkogluTVF12},
or a mixture of both types of features \citep{DBLP:journals/datamine/OteyGP06} in addition to one-class classification-based approaches \cite{conf/icmla/JanssensFP09,conf/incdm/PauwelsA11}.

We refer the reader to a comprehensive survey on outlier detection for more discussion and details \citep{ChandolaBK12}
as well as a recent book by \citep{aggarwal2013outlier} on outlier analysis with comprehensive details on these techniques.


\vspace{0.15in}
\noindent \textbf{Anomalies in Static Graph Data}
\vspace{0.1in}

\begin{wrapfigure}[7]{r}{0.39\textwidth}
\vspace{-0.4in}
\includegraphics[width=135pt]{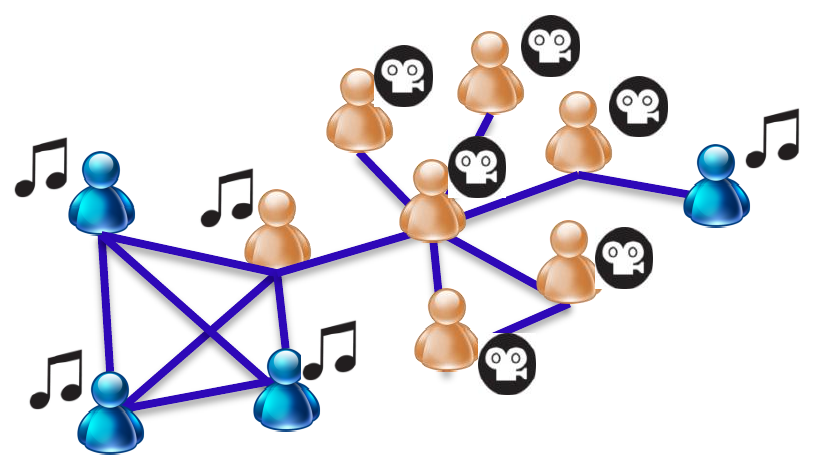}
\vspace{-1.2in}
\end{wrapfigure}
We will study anomaly detection in graph data under two settings: 1) plain graphs, and 2) attributed graphs.
An attributed graph is a graph where nodes and/or edges have features associated with them.
For example in a social network, users may have various interests, work/live at different locations, be of various education levels, etc.
while the relational links may have various strengths, types, frequency, etc.
A plain graph, on the other hand, consists of only nodes and edges among those nodes, i.e. the graph structure.

While the specific definition of the graph anomalies may vary, a general definition for the anomaly detection problem for static graphs can be stated as follows:

\begin{definition}[Static-Graph Anomaly Detection Problem] \\
\indent \textbf{Given} the snapshot of a (plain or attributed) graph database, \\
\indent \textbf{Find} the nodes and/or edges and/or substructures that are ``few and different'' or 
\indent \indent deviate significantly from the patterns observed in the graph.
\end{definition}

\subsection{Anomalies in static plain graphs}
\label{sec:staticplain}

For a given plain graph, the only information about it is its structure. This category of anomaly detection methods thus exploit the structure of the graph to find patterns and spot anomalies. These structural patterns can be grouped further into two categories: {\em structure-based} patterns and {\em community-based} patterns.


\subsubsection{Structure based methods}

\noindent
We organize the structure-based approaches into two: feature-based and proximity-based.
The first group exploits the graph structure to extract graph-centric features such as node degree and subgraph centrality, while the second group uses the graph structure to quantify the closeness of nodes in the graph to identify associations.

\vspace{0.15in}
\noindent \textbf{Feature-based approaches:}
\vspace{0.1in}

\noindent \emph{Main idea:} This group of approaches uses the graph representation to extract structural graph-centric features that are sometimes used together with other features extracted from additional information sources for outlier detection in the constructed feature space. Essentially, these methods transform the graph anomaly detection problem to the well-known and understood outlier detection problem.

\vspace{0.05in}
\noindent \emph{Graph-centric features:}
One could use the given graph structure to compute various measures associated with the nodes, dyads, triads, egonets, communities, as well as the global graph structure \citep{HendersonEFALMPT10}.
These features have been used in several anomaly detection applications including Web spam \citep{becchetti2006link} and network intrusion \citep{conf/kdd/DingKBKC12} as we will discuss in detail in Section \ref{sec:fraud}.

The {\em node-level} features include (in/out) degrees, centrality measures such as eigenvector \citep{citeulike1143909}, closeness \citep{PhysRevLett92118701}, and betweenness \citep{freeman77} centralities, local clustering coefficient \citep{Watts1998}, radius \citep{KangHadi11}, degree assortativity, and most recently, roles \citep{HendersonGETBAKFL12}.
The {\em dyadic} features include reciprocity \citep{Akoglu2011reciprocity}, edge betweenness,  number of common neighbors, as well as several other local network overlap measures \citep{libenNowell2003theLinkPrediction,conf/websci/GupteE12}.
\citep{Akoglu2009oddball} introduce {\em egonet} features such as its number of triangles, total weight, principal eigenvalue, etc. as well as their pairwise correlation patterns.
\citep{conf/kdd/HendersonGLAETF11} enrich and extend the possible graph-based features with recursively aggregating existing features.
The {\em node-group-level} features can be listed as compactness measures, such as density, modularity \citep{Newman06062006}, and conductance \citep{andersen2006local}.
 Finally, examples to {\em global} measures include number of connected components, distribution of component sizes \citep{KangMAF10}, principal eigenvalue, minimum spanning tree weight, average node degree, global clustering coefficient, to name but a few.

\vspace{0.05in}
\noindent \emph{Approaches:}
A feature-based anomaly detection technique called \textsc{OddBall} is proposed by \citep{Akoglu2009oddball}, which extracts egonet-based features and finds patterns that most of the egonets of the graph follow with respect to those features. As such, this method can spot anomalous egonets (and hence anomalous nodes), as those that do not follow the observed patterns.

\begin{wrapfigure}[6]{r}{0.23\textwidth}
\vspace{-0.3in}
\hspace{-0.25in} \includegraphics[width=90pt]{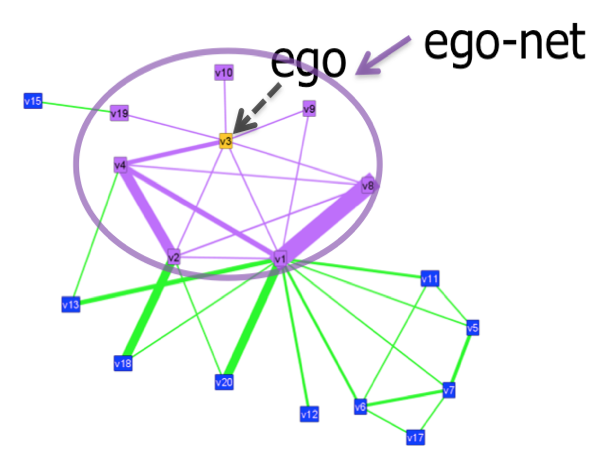}
\label{fig:egonet}
\end{wrapfigure}
An {\em egonet} is defined as the 1-step neighborhood around a node; including the node, its direct neighbors, and all the connections among these nodes (an example is shown on the right figure). More formally an egonet is the induced 1-step sub-graph for each node. Given the egonets, the main question and challenge is which features to look at, as there is a long list of possible graph-based measures that can be extracted as egonet features. The paper proposes a carefully chosen subset of features (e.g. number of triangles, total weight of edges, etc.) that are (1) observed to yield patterns across a wide range of real-world graphs, and (2) fast to compute and easy to interpret.

The egonet features are then studied in pairs and several patterns in the form of power-laws are observed among strongly related features (e.g. number of neighbors and number of triangles). For a given egonet, its deviation from a particular pattern is computed based on its ``distance'' to the relevant power-law distribution.
Each egonet then receives a separate deviation, or outlierness, score with respect to each pattern.

The multiple scores a node receives from various observed patterns brings up the question of how to combine them to obtain the final scores or final ranking.
Several works \citep{conf/icdm/GaoT06,conf/kdd/LazarevicK05} have proposed solutions to how to unite multiple outlierness scores.
This problem is addressed in works on outlier ensembles, as discussed in \cite{EnsembleCharu,Zimek2014EUO}.

There are several advantages of analyzing the egonet features in pairs, rather than in union. First, this facilitates the visualization of the patterns and outliers in 2-d for post-analysis. Second, the low dimensionality of the feature space helps with interpretability of the results, that is, one can tell  what type
of anomalies a node belongs to based on its deviation from a particular pattern, or ``law''. 
As an example, in \citep{KangLKF14}, the authors propose a package for visualization of billion-scale graphs by focusing on correlation plots (node features in pairs), as well as the spy plot and distribution plots for various features. The visualization tool is carefully designed to make the outliers pronounced even by a simple inspection.

Later work by \citep{conf/kdd/HendersonGLAETF11} extends the feature base by recursively combining node-based (``local'') and egonet-based (neighborhood) features.
A recursive feature is defined as some aggregate value (e.g. mean, min, max) computed over any existing feature value (including recursive ones)  among a node's neighbors.
Intuitively, local and egonet features capture neighborhood information, whereas recursive features enable to go beyond direct neighborhood to capture more of ``regional'' or behavioral information. An iterative procedure with run time complexity linear in graph size is detailed in the paper to compute recursive features and prune highly correlated features on the go.

\vspace{0.15in}
\noindent \textbf{Proximity-based approaches:}
\vspace{0.1in}

\noindent \emph{Main idea:}
This group of techniques exploits the graph structure to measure closeness (or proximity) of objects in the graph. These methods capture the simple auto-correlation between
these objects, where close-by objects are considered to be likely to belong to the same class (e.g., malicious/benign or infected/healthy).

\vspace{0.05in}
\noindent \emph{Approaches:}
Measuring the importance of the nodes in a graph is one of the most widely studied graph problems.
PageRank \citep{BrinP98}  is one of the most popular algorithms which
is based on random walks.
A random walk on the (unweighted) graph jumps randomly from node to node.
If currently present on a node $u$, a random
walk in the next step jumps to one of its neighbors with equal probability $1/d_u$
 where $d_u$ is the degree of node $u$.
 The stationary probability distribution of the random walk
 on the graph is then considered to rank the nodes by their ``importance''.

 This walk is known to converge if the transition matrix, the entries of which denote the jump probabilities between neighboring nodes, is stochastic, aperiodic, and irreducible \citep{feller1968introduction}.
 On an undirected
 graph, the stationary probability of a random walk at
 node $u$ is directly proportional to its degree $d_u$, and is
 independent of the starting node. On directed graphs,  it is probable that the irreducibility condition, which states that there is a non-zero
 probability of going from any one
 node to any other, will be unmet (e.g., in the existence of sink nodes and multiple strongly connected components).
 To resolve these issues, a random restart of the walk is performed with a certain probability $\alpha \in (0,1)$ (a.k.a. the damping factor),
 where the restart node is chosen at random.

 A widely used graph-closeness measure that is also based on random walks but with an extension of
 restarts to a particular node is the Personalized PageRank (PPR) \citep{haveliwala2003topicsensitive}.
 Given a restart node $q$ and the parameter
 $\alpha$ consider the random
 walk with restart,
 starting at node $q$, such that at any step
 when currently
 present at a node $u$, it chooses any of its neighbors with equal probability $(1-\alpha) /d_u$,
 and returns to the restart node $q$ with probability $\alpha$.
 The stationary probability at any node $v$ of the random walk with restart
  is defined as the
PPR score of $v$ with respect to the
 restart node $q$.
 A more general version of this measure can be given for a set $Q$ of restart nodes,
 where their restart probabilities sum to $\alpha$. This type of PageRank computation is often referred to as the Biased PageRank.
The stationary distribution of probabilities indicates the proximity (or closeness) of each node in the graph with respect to the (set of) restart node(s), and is higher for the nodes that have many, short, and high weighted paths to the restart node(s).

Another graph proximity measure that quantifies the closeness of two nodes in the graph is SimRank \citep{SimRank02}, which computes similarity
of the structural context in which the graph objects occur, based
on their relationships with other objects. It is often thought as measuring how soon two random surfers starting
from the two nodes are expected to meet each other by
randomly walking ``backwards'' in the graph. Several variants of SimRank are also proposed by \citep{Antonellis2008simrank,conf/cikm/ZhaoHS09,ascos13}.

Finally, many link prediction approaches essentially quantify the similarity or closeness of pairs of nodes in the graph. Several such measures of varying computational complexity exist. The simple ones include the Jaccard proximity, which is the normalized number of common neighbors of the two nodes. Others include the total number of paths or node-disjoint paths. The slightly more complex Katz measure \cite{katzb145} counts all the paths weighted inversely proportional to the path length.
For a well documented list and evaluation of these as well as other measures, we refer the reader to \cite{libenNowell2003theLinkPrediction}.


\subsubsection{Community based methods}

\noindent \emph{Main idea:}
The cluster or community-based methods for graph anomaly detection rely on finding densely connected groups of ``close-by'' nodes in the graph and spot nodes and/or edges that have connections across communities. In fact, the definition of anomaly under this setting can be thought of as finding ``bridge'' nodes/edges that do not directly belong to one particular community.

\vspace{0.05in}
\noindent \emph{Approaches:}
Methods that exploit communities or proximity of nodes in the graph to spot (node) anomalies in bipartite graphs include \citep{conf/icdm/SunQCF05}.
Several real-world data can be represented with bipartite graphs where the bridge nodes reveal interesting phenomena. Examples include
publication networks: authors vs. (unusual) papers written by authors from different research communities; P2P networks: users vs. (cross-border) files; financial trading networks: stocks vs. (cross-sector) traders; and customer-product networks: users vs. (``cross-border'') products.

The two main problems addressed in \citep{conf/icdm/SunQCF05} are (P1) how to find the community of a given node, which is also referred as the ``neighborhood'' of a node, and
(P2) how to quantify the level of the given node to be a bridge node.
 For (P1), the authors use random-walk-with-restart-based Personalized PageRank (PPR) scores \citep{haveliwala2003topicsensitive} of all the nodes with respect to the given node, where those nodes with high PPR scores constitute the neighborhood of a node. On similar lines, for (P2) the pairwise PPR scores among all the neighbors of the given node are aggregated by averaging to compute a so-called ``normality'' score of a node. Intuitively, nodes with low normality scores have neighbors with low pairwise proximity to one another. This suggests that the neighbors lie in different, separate communities, which makes the given node resemble a bridging node across communities.

\textsc{Autopart} \citep{Chakrabarti04} is based on the notion that nodes with similar neighbors are clustered together, and the edges that do not belong to any structure constitute anomalies (e.g. cross-cluster bridge edges). Similarly, nodes that have many cross-connections to multiple different communities are considered not to belong to any particular cluster and thus also constitute anomalies.
For finding communities in a graph, the algorithm re-organizes the rows and columns of the adjacency matrix into a few homogeneous blocks (of either low or high density). These blocks have the property of containing nodes that are more densely connected together than with the rest of the nodes in the graph---which is the underlying idea in clustering.  \citep{Chakrabarti04} develops a parameter-free, iterative algorithms based on the Minimum Description Length principle \citep{rissanen99} for rearranging the rows and columns, as well as for finding the best number of blocks or node groups automatically without requiring any user input.

Another method that aims to spot (node and edge) anomalies based on graph communities \citep{Tong@SDM11} relies on matrix factorization.
Matrix factorization has been used to address several problems ranging from dimensionality reduction \citep{DBLP:journals/jmlr/NikulinH12,DBLP:journals/ieicet/AmbaiUY11} to (graph) clustering \citep{DBLP:journals/tkde/WangRDD12,DBLP:conf/sdm/KuangPD12}.
The factorization of a data matrix $\mat A$ is often formulated as $\mat{A} = \mat{X}\times \mat{Y} + \mat{R}$, where $\mat X$ and $\mat Y$ are the low rank factors and $\mat R$ denotes the residual matrix.
In traditional non-negative matrix factorization (\textsc{NMF}),  there exists additional constraints on the non-negativity of both $\mat X$ and $\mat Y$, which for example aids in determining the communities.
Different from this traditional approach, the main idea for finding anomalies is to waive these original constraints but instead enforce non-negativity constraints on the {\em residual} matrix for interpretability (hence the name \textsc{NrMF}).
The approach proves effective in spotting ``strange'' connections, such as port-scanning-like or ddos-like activity, bridging connections, as well as bipartite-core structures with the help of the non-negative residual matrix.

The ``bridge'' nodes and/or edges can be seen as intrusive connectors and/or connections that cross the community boundaries in computer security. For example, \citep{conf/kdd/DingKBKC12} regards intrusion as entering a community to which one does not belong, and looks for communication that does not respect the community boundaries. Analysis shows that cut-vertices (vertices the removal of which disconnects the graph into components) correspond well with ground-truth traffic sources that attempted an intrusion, by sending malicious or unwanted traffic.
This work essentially shows one of the real-world applications that community-based anomaly detection methods prove to be effective.

Other community-based network outlier detection methods directly focus on network clustering, and in the process, spot hubs and outliers as a by-product \citep{SCAN1281280,conf/icdm/SunHHDZF10}.
To find network clusters, \textsc{SCAN} \citep{SCAN1281280} exploits the neighborhood of vertices; vertices sharing many neighbors are grouped into the same clusters. As such, vertices that are bridging many clusters are labeled as hubs, whereas those that cannot be assigned to any community are flagged as outliers.
To overcome the issue of selecting the minimum similarity threshold parameter of \citep{SCAN1281280},
 \citep{conf/icdm/SunHHDZF10} proposes a novel clustering framework called \textsc{gSkeletonClu} that also aims to find hubs and outliers as byproduct of the graph clustering.

\subsection{Anomalies in static attributed graphs}
\label{sec:staticattributed}

For certain kinds of data, it is possible to have a richer graph representation, in which nodes and edges exhibit (non-unique) attributes. Examples to such graphs include social networks with user interests as attributes, transaction networks with time, location, and amount as attributes, cargo shipments with visited ports, financial information, type of transported goods as attributes, and so on.\footnote{We will use the words `attribute' and `feature' interchangeably throughout text.}

This category of anomaly detection methods on attributed graphs exploit the structure as well as the coherence of attributes of the graph to find patterns and spot anomalies. These methods can also be grouped into two: {\em structure-based} and {\em community-based} methods. In a nutshell, the structure-based methods exploit frequent substructure and subgraph patterns to spot deformations in these pattens, while community-based methods aim to spot what is called community-outliers that do not exhibit the same characteristics as the others in the same community.



\subsubsection{Structure based methods}

\noindent \emph{Main idea:}
Structure based approaches mainly aim to identify substructures in the graph that are rare structurally, i.e. connectivity-wise, as well as attribute-wise. As such, inverse of frequent attributed subgraphs are sought out. The differences from these normative substructures are quantified in various ways as we describe below.

\vspace{0.05in}
\noindent \emph{Approaches:}
One of the earliest works on attributed graph anomaly detection by \citep{NobleC03} addresses two related problems: (P1) the problem of finding unusual substructures in a given graph, and (P2) the problem of finding the unusual subgraphs among a given set of subgraphs, in which nodes and edges contain (non-unique) attributes.
Main insight to solve these problems is to look for structures that occur infrequently, which are roughly opposite to what is called the ``best substructures''.
 Intuitively, best substructures are those that occur frequently in the graph and thus can compress the graph well. An information-theoretic formulation based on the Minimum Description Length (MDL) principle \citep{rissanen99} that trades off between compression quality and the size of such substructures (as the entire graph is the best compressor) is devised as an objective.

The main idea for detecting unusual substructures (P1) is to define a measure that is inversely related to the MDL-based measure defined for the best substructures and rank substructures by this new measure.
Similarly, the main idea for finding the unusual subgraphs (P2) is to define a measure that penalizes those subgraphs containing few common (i.e. best) substructures, making them more anomalous.

The methods by \citep{NobleC03} essentially build on frequent subgraphs with categorical attributes.
On the other hand, most often datasets come with a mix of both numerical and categorical attributes, e.g. dollar amounts in transaction data and number of (e.g., Ping, SYN, etc.) requests in network log data.
Treating each numerical value as a distinct attribute loses ordering and closeness information.
To address this problem \citep{conf/cikm/DavisLMR11} proposed {\em discretizing} the numerical attributes, where the majority ``normal'' values are assigned the same single categorical attribute, and all other values are assigned their ``outlierness'' score. Several discretization mechanisms, e.g. based on fitting probability density functions, $k$-NNs, outlier detection (in particular LOF \citep{conf/sigmod/BreunigKNS00}), and clustering (CbLOF \citep{journals/prl/HeXD03}), have been studied.
We also include other discretization techniques that could apply under this setting such as SAX \citep{conf/dmkd/LinKLC03}, MDL-binning \citep{journals/jmlr/KontkanenM07}, and minimum entropy discretization \citep{conf/ijcai/FayyadI93}.

Later work by \citep{conf/icdm/EberleH07} follows a different insight to look for anomalies than the previous work. Rather than focusing on infrequent substructures, they go after those substructures that are {\em very similar to, though not the same as, a normative (i.e. best) substructure}.
A statement by United Nations Office on Drugs and Crime corroborates this insight:
``The more successful money-laundering apparatus is in imitating the patterns and behavior of legitimate transactions, the less the likelihood of it being exposed.''

Using the insight that an intruder would make at most a certain number of changes to blend in with the  normal data instances and lower their chances of being detected glaringly, the work by \citep{conf/icdm/EberleH07} formulates three types of anomalous cases based on modification, insertion, and deletion.
They formulate various anomaly scores that use both (in)frequency and modification cost (the lower, the more anomalous).
We note that the anomalies are assumed to consist of only one type of anomaly, which is prone to missing e.g., a deletion followed by a modification.

On similar lines, \citep{conf/sdm/LiuYYHY05} use subgraphs of attributed graphs for detecting non-crashing software bugs.
In this type of application domain, every execution of a software program is represented as an attributed graph called behavior graph, where nodes denote functions (attributed with function names), and (directed) edges depict function calls or function transitions.  Different from previous methods discussed so far, the idea here is to train a classification model that takes as input
positive and negative behavior graphs for correct and incorrect executions, respectively.
First, (closed) frequent subgraphs are extracted from a set of behavior graphs, which are then used as features in training a classification model.

The pattern-based (e.g. frequent substructures) anomaly detection techniques as described above make them interpretable and amenable for post-analysis by domain experts to reveal the root cause.
Moreover, these methods are quite generally defined such that they can be applied on various types of data and scenarios where the data can be represented as attributed (sub)graphs (like the software execution flow-graphs). On the other hand, this generality comes at a cost of high false positive rates, as not all rare occurrences can be attributed to anomalous cases.
Furthermore several user-specified thresholds, such as the amount of alteration threshold or subgraph frequency threshold, make it hard to trade off false positive and false negative rates by the user.



\subsubsection{Community based methods}

\noindent \emph{Main idea:}
These approaches aim to identify those nodes in a graph, often called the community outliers, the attribute values of which deviate significantly from the other members of the specific communities that they belong to. 
For example, a smoker in a community of vastly non-smoker baseball players is an example of a community outlier.
As such, communities are analyzed based on both link and attribute similarities of the nodes they consist of. 
While some methods aim to detect outliers simultaneously with detecting the communities in the graph, some perform the outlier detection as a second step after performing the attributed graph clustering.

\vspace{0.05in}
\noindent \emph{Approaches:}
\citep{conf/kdd/GaoLFWSH10} differentiates graph-based community outlier detection from three closely related problems; namely, global outlier detection that only considers node attributes,
structural outlier detection that only considers links (e.g. \citep{SCAN1281280}) (as is discussed in the previous section), and local outlier detection that only considers attribute values of direct neighbors.
While interesting on their own right, these three types of methods are prone to miss outliers in the unison of these---outliers with respect to other community members' attributes.
They develop a unified probabilistic model that simultaneously finds communities as well as spot community outliers. The unsupervised learning algorithm called \textsc{CODA} alternates between the two steps of parameter estimation (fixed cluster assignment), and inference for cluster assignments (fixed parameters). As with the nature of such learning algorithms, the good initialization of clusters at the beginning is a crucial step for the algorithm to reach a good solution. Moreover, the convergence of the algorithm is not guaranteed. One way that is used to find a good initialization is to employ a graph clustering algorithm to find a first-cut good quality clustering based on only the link structure, which also helps with faster convergence.

Recently \citep{Muller13} developed a node outlier ranking technique in attributed graphs called \textsc{gOutRank}. Different from \citep{conf/kdd/GaoLFWSH10}, their main insight into community outlier detection is the fact that the complex anomalies could be revealed in only a subset of relevant attributes (a.k.a. subspaces). This becomes more apparent especially in high dimensional feature spaces due to the curse of dimensionality \citep{Beyer99whenis}.
Roughly speaking, all objects appear to be sparse and dissimilar in high dimensions, or in other words, all the distances between pairs of objects look similar causing all the objects to be equally (dis)similar to one another.
In their work, they also consider quantifying the degree of deviation for each node-outlier which is beyond binary detection.
As such, they address two main challenges associated with community outlier detection in attributed graphs; the selection of subgraphs and subspaces, and the scoring of nodes in multiple subspace clusters.

Other related work mainly addresses the problem of attributed graph clustering without focusing on outlier detection, including \citep{conf/kdd/BodenGHS12,journals/datamine/GunnemannBS12,BGS12,conf/sdm/AkogluTMF12,conf/icdm/GunnemannFBS10}. These methods could form the basis for community outlier detection in a post-processing step, as opposed to integrated clustering and outlier detection in one algorithm as with the techniques discussed above. During post-processing, nodes that could not be assigned to a ``large enough'' community (e.g., singletons or micro-clusters) could be analyzed further, or the nodes the removal from a community of which increases a ``fitness'' score of the community can be flagged as abnormal.

\subsubsection{Relational learning based methods}

\noindent \emph{Main idea:}
This group consists of network-based collective classification algorithms the main idea of which is to exploit the relationships between the objects to assign them into classes, where the number of classes is often two: anomalous and normal.
Different from proximity-based approaches which aim to quantify auto-correlations among graph objects, these algorithms are often more complex and thus can model and exploit more complex correlations between the graph objects.

\vspace{0.05in}
\noindent \emph{Approaches:}
Classification is the problem of assigning class labels to, or shortly labeling, data instances based on their observed attributes.
Anomaly detection can be formulated as a classification problem, when one has a representative labeled data available.
For example, determining whether a Web page is spam or non-spam based upon the words that
appear in it and identification of benign/malicious web pages, fraud/legitimate transactions, etc. can all be thought of as two-way classification problems.
When the labeled data size is reasonably large, one can employ fully supervised classification, where the labeled data is used for model learning. When labeled data is scarce, but still available, one can employ semi-supervised classification, where the learning is done by simultaneously using labeled and unlabeled data. 

In traditional statistical machine learning approaches, the instances are often assumed
to be independent, and identically distributed (i.i.d) and often the learning algorithms ignore the dependencies among data instances.
Relational classification, on the other hand,
is the task of inferring the class labels of a network of objects simultaneously or collectively. The underlying
assumption in relational classification is that the relationships between objects carry important
information for classifying the objects, such as two linked Web pages. In many cases, there is a simple auto-correlation between
the objects, where the linked objects are likely to have the same labels (e.g. spam pages link to other spam pages, infected people are linked to other infected people). In other cases,
more complex correlations may be exhibited (e.g. fraudsters trade with honest people and not with other fraudsters).

There exist a large amount of research on relational classification methods \citep{macsk03,TaskerPK02,FriedmanGKP99,getoor2003link,JensenNG04,neville00iterative,Neville03Learning,nevilleJensen2003b}. Generally, these methods exploit one or more of the following input:

\vspace{-0.05in}
\begin{enumerate}
\item the class labels of its neighbors, and
\item the node attributes (features),
\item the attributes of the node's neighbors.
\end{enumerate}
\vspace{-0.05in}

We note that although it is possible that some methods described in this section are amenable to use only the first type of  information, i.e. nodes' class labels, and need not exploit node attributes, most methods are easily generalizable to incorporating node attribute information, if available. Thus, we cover these methods in this section that is attributed to anomaly detection in attributed graphs, and remark that some methods do apply to plain graphs as well.

Relational classification methods can be categorized into local and global methods \citep{SenNBGGE08}. The local  algorithms build local predictive models for the class of a node in the network and  use often iterative inference procedures to
collectively classify the unlabeled objects.
The second group of algorithms define a global formulation of class dependencies and use inference algorithms to solve for the assignments that would maximize the joint probability distribution.

The techniques for the local methods can differ in both the local models and the
inference methods that they use.
\citep{Chakrabarti07WWW} use Naive Bayes models for the local attributes of
the object and the class labels of the neighbor objects. They then use mean field relaxation labeling
for the inference. \citep{neville00iterative} also use a Naive Bayes model for the attributes,
but they use an iterative classification algorithm (ICA) for inference. In later work, they investigate the use of
relational dependency networks (RDNs) and the inference algorithm is based on Gibbs sampling \citep{nevilleJensen2003b}.
\citep{getoor2003link} use logistic regression as a local model and ICA for inference but they explore various ways of aggregation that can be used for the class labels of the related objects. For sparsely labeled networks, \citep{conf/kdd/GallagherTEF08} propose ways to infer ``ghost'' edges based on graph closeness to improve classification performance.

As for the global methods, \citep{FriedmanGKP99} use
probabilistic relational models (PRMs) as a (full joint) model and then use Loopy Belief Propagation (LBP) \citep{Yedidia2003} for the
inference. \citep{TaskerPK02} use relational Markov networks (RMNs) as a (full joint) model and also use
LBP for inference. \citep{macsk03} propose a simple baseline algorithm called (probabilistic) weighted-vote relational network  (wv-RN) classifier  where they use only the
class labels of objects for classification; they infer the class label of an object by taking a weighted
average of the potentially inferred class labels of the related objects iteratively. Other global formulations are based on Markov logic networks (MLNs).

All in all, the relational inference algorithms mentioned above can be listed as

\vspace{-0.05in}
\begin{itemize}
\item Iterative Classification Algorithm (ICA)
\item Gibbs Sampling
\item Loopy Belief Propagation
\item Weighted-Vote Relational Network Classifier
\end{itemize}
\vspace{-0.025in}

All these algorithms are fast, iterative, approximate inference algorithms, since exact inference is known to be NP-hard in arbitrary networks \citep{cooper90complexity}. Moreover, convergence is not guaranteed for any of them.
Node ordering for updates (e.g., random, diversity-based) may alter the classification results.
For local methods, additional challenges include feature construction and local classification. For feature construction one has to decide whether to consider in-, out-, or both neighbors, and aggregation method of neighbor labels (e.g., max, mode, count), as well as choice of neighbors to consider (e.g., all, top-k most confidently labeled). With respect to local classification, one requires training data, and has to choose the classifier type (e.g., Naive Bayes, logistic regression, k-NN, SVM).

With respect to scalability, these methods mostly rely on message passing or information aggregation over neighbors and thus scale linearly with number of edges in the graph. Recently, techniques to speed up inference for massive graphs, especially based on LBP, have been proposed by \citep{KangCF11,KoutraKKCPF11}.

Before concluding this section on graph anomaly detection techniques in static graphs, we provide a summary and qualitative comparison of the detection algorithms presented in this section in Table~\ref{tab:static_qualitative}.

\begin{landscape}
 \begin{table}[tbh]
 \centering
 \caption{Qualitative and quantitative comparison of anomaly detection algorithms for \textit{static} graphs. The first four columns refer to the type of graphs that an algorithm can be applied to (with or without weights on the edges, with or without attributes for the nodes);
 ``Linear'' holds true for those methods that have time complexity linear in the number of edges of the input graph (and more otherwise);
 ``Parameter-free'' methods correspond to those that do not expect any user-specified input parameters;
 ``Output format'' corresponds to the output type/format of the method (e.g. anomaly scores and their ranges, binary output/classification e.g. anomalous or not);
and ``Visualization'' refers to the graphical means used --if any-- to present the anomalous instances to the user (e.g., distribution plots, graph with the anomalous nodes/edges annotated). }
\label{tab:static_qualitative}
\scriptsize{
\begin{tabular}{|l||c|c|c|c||c||c||c||c|} \hline
\multicolumn{1}{|g||}{} & \multicolumn{4}{|g||}{\textbf{Graphs}} &  \multicolumn{1}{|g||}{} & \multicolumn{1}{|g||}{} & \multicolumn{1}{|g||}{} & \multicolumn{1}{|g|}{} \\ \cline{2-5}
\multicolumn{1}{|g||}{\textbf{Algorithm}} & 
\multicolumn{1}{|g|}{{\begin{sideways}\textbf{Weighted} \end{sideways}}} & 
\multicolumn{1}{|g|}{{\begin{sideways}\textbf{Unweighted} \end{sideways}}} & 
\multicolumn{1}{|g|}{{\begin{sideways}\textbf{Attributed} \end{sideways}}} & 
\multicolumn{1}{|g||}{{\begin{sideways}\textbf{Plain} \end{sideways}}} &
\multicolumn{1}{|g||}{{\begin{sideways}\textbf{Linear} \end{sideways}}} & 
\multicolumn{1}{|g||}{{\begin{sideways}\textbf{Parameter-free} \end{sideways}}} & 
\multicolumn{1}{|g||}{{\begin{sideways}\textbf{Output format} \end{sideways}}} & 
\multicolumn{1}{|g|}{{\begin{sideways}\textbf{Visualization } \end{sideways}}}\\ \hline \hline
\multicolumn{1}{|g||}{\textsc{OddBall} \citep{Akoglu2009oddball}, \citep{conf/kdd/HendersonGLAETF11}} & \cmark & \cmark & \xmark & \cmark & \xmark & \cmark &  $[0,\infty]$ node anomaly scores & pairwise feature scatter plots, egonets \\ \hline 
\multicolumn{1}{|g||}{\citep{conf/icdm/SunQCF05}}                    & \cmark & \cmark & \xmark & \cmark & \xmark & \cmark &  $[0,1]$ node normality scores & score distribution \\ \hline 
\multicolumn{1}{|g||}{\textsc{Autopart} \citep{Chakrabarti04}}      & \xmark & \cmark & \xmark & \cmark & \cmark & \cmark &  binary edge classification & adjacency matrix organized by node clusters \\ \hline
\multicolumn{1}{|g||}{\textsc{NNrMF} \citep{Tong@SDM11}} & \cmark & \cmark & \xmark & \cmark & \cmark & \cmark &  binary edge/node classification & residual matrix \\ \hline 
\multicolumn{1}{|g||}{\citep{conf/kdd/DingKBKC12}}                  & \cmark & \cmark & \xmark & \cmark & \xmark & \cmark &  binary node classification & egonets \\ \hline 
\multicolumn{1}{|g||}{\textsc{SCAN} \citep{SCAN1281280}}          & \xmark & \cmark & \xmark & \cmark & \cmark & \xmark &  binary node classification & clustering with hub\&outlier nodes \\ \hline 
\multicolumn{1}{|g||}{\textsc{gSkeletonClu} \citep{conf/icdm/SunHHDZF10}}          & \cmark & \cmark & \xmark & \cmark & \xmark & \cmark &  binary node classification & clustering with hub\&outlier nodes \\ \hline
\multicolumn{1}{|g||}{\textsc{Subdue} \citep{NobleC03}}          & \xmark & \cmark & \cmark & \xmark & \xmark & \xmark &  substructure anomaly score $\in \mathbb{N}$ & graph substructures \\ \hline
\multicolumn{1}{|g||}{\textsc{Subdue} \citep{NobleC03}}          & \xmark & \cmark & \cmark & \xmark & \xmark & \xmark &  $[0,1]$ subgraph anomaly score & subgraphs \\ \hline
\multicolumn{1}{|g||}{\textsc{Subdue} \citep{conf/icdm/EberleH07}}          & \xmark & \cmark & \cmark & \xmark & \xmark & \xmark &  $[0,\infty]$ subgraph anomaly score & modified subgraphs \\ \hline
\multicolumn{1}{|g||}{\citep{conf/sdm/LiuYYHY05}}          & \xmark & \cmark & \cmark & \xmark & \xmark & \xmark &  binary graph classification & graphs with traced-back crashing points \\ \hline
\multicolumn{1}{|g||}{\textsc{CODA} \citep{conf/kdd/GaoLFWSH10}}          & \cmark & \cmark & \cmark & \xmark & \cmark & \xmark &  binary node classification &  graph clustering with community outlier nodes\\ \hline

\multicolumn{1}{|g||}{\textsc{gOutRank} \citep{Muller13}}          & \xmark & \cmark & \cmark & \xmark & \cmark & \xmark & $[0,\infty]$ node anomaly scores  &  subspace clustering and outlier nodes\\ \hline

\end{tabular}
}
 \end{table}
 \end{landscape}

\section{Anomaly Detection in Dynamic Graphs}
\label{sec:dynamic}

\subsection{Overview: Event detection in time series of data points}
In the literature, there is abundance of work on event detection on data series: statistical quality control \citep{Montgomery97}; the famous auto-regressive moving average model used for predictions  \citep{BoxJ90}; a drift detection method \citep{GamaMCR04}; a chart-based approach for monitoring temporal, medical data \citep{GriggFS03}; change detection in categorical data \citep{BayP99}; StreamKrimp, an MDL-based algorithm \citep{LeeuwenS08}; detection of disease outbreaks \citep{WongMSW05}. A nice tutorial that covers event detection in data series is \citep{NeillW09} and a survey on outlier detection for temporal data is \citep{GuptaGAH13}.


\subsection{Event detection in time series of \emph{graph} data}
\label{sec:timeEvolvingGraphs}

\begin{figure}[h!]
\vspace{-0.2in}
\centering
   \includegraphics[width=0.85\textwidth]{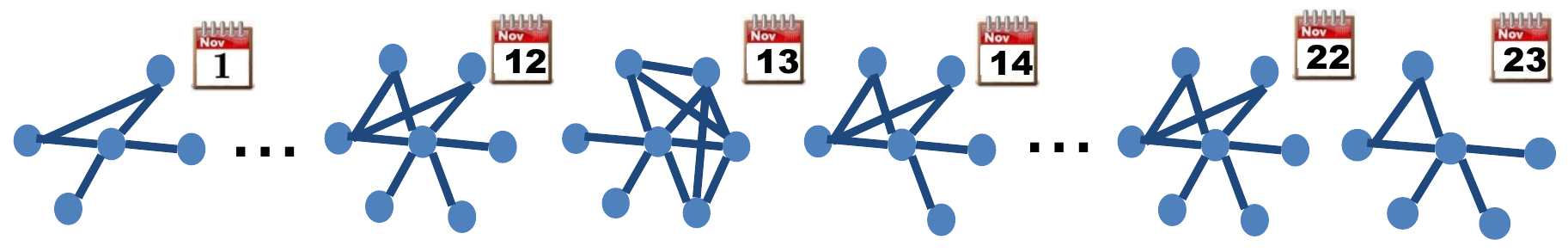}
\label{fig:dynamic}
\vspace{-0.1in}
\end{figure}

This section provides an overview of the anomaly detection algorithms that have been proposed for \emph{dynamic} or \emph{ time-evolving} graphs (i.e. sequences of static graphs), the evolution of which as well as their communities have been studied by several research groups \citep{LeskovecKF05, BackstromHKL06}. In addition, \citep{CharuEvolSurvey14} provides a comprehensive survey on evolutionary network analysis. The anomaly detection problem for dynamic graphs, which is the main focus of our survey, is also known as temporal anomalous pattern detection, event detection, change-point detection, and is commonly defined as follows:

\begin{definition}[Dynamic-Graph Anomaly Detection Problem] \\
\indent \textbf{Given} a sequence of (plain or attributed) graphs,  \\
\indent \textbf{Find} (i) the timestamps that correspond to a \emph{change} or \emph{event}, as well as \\
\indent \indent $\;\;$ (ii) the top-$k$ nodes, edges, or parts of the graphs that contribute most to the \\
\indent \indent $\;\;\;\;\;\;\;$ change (\emph{attribution}). 
\end{definition}

Depending on the application domain, the requirements of the algorithms vary, but among the most usual desired properties are:
\begin{itemize*}
\item {\em Scalability}. As instructed by the size and volume of the graphs that are produced daily, ideally, the algorithms should be linear or sub-linear on the size of the input graphs. In the dynamic setting, an additional, desired property is that the algorithm should be linear on the size of the {\em update} of the input graphs.
\item {\em Sensitivity to structural and contextual changes}. The anomaly detection methods should be able to discern structural differences between the input graphs under comparison (e.g., missing/new edges, missing/new nodes, changes in the weights of the edges), as well as changes in other properties of the graphs, such as labels of the nodes or edges.
\item {\em Importance-of-change awareness}. The algorithms should be sensible to the type and extent of change. Changes in ``important'' nodes, edges or other graph attributes should result in greater anomaly scores, than changes in less important structures.
\end{itemize*}

A brief overview of the anomaly detection algorithms for time-evolving graphs is given in \citep{BilginY08}. However, the abundance of time-evolving graphs in the recent years has led to increasing interest in them, and subsequently new research has been carried out in this area. In the following subsections, we classify the dynamic graph anomaly detection algorithms based on the type of ``graph summary'' or ``footprint'' they use, and the type of \emph{events} they detect: (i) feature-based (e.g. nodes, edges, edge weights), (ii) decomposition-based, (iii) community or clustering-based, and (iv) window-based.

\subsubsection{Feature-based Events}
\label{sec:dynamicfeatureextraction}

\noindent \emph{Main idea:} The key idea behind the feature-based methods is that similar graphs probably share certain properties, such as degree distribution, diameter, eigenvalues\citep{centralities} \citep{graph_statistics}.
The general approach in detecting anomalous timestamps in the evolution of dynamic graphs can be summarized in the following steps:
\begin{itemize*}
\item Extract a ``good summary'' from each snapshot of the input graph.
\item Compare consecutive graphs using a distance --or equivalently, similarity-- function. A nice survey on similarity measures is given in \citep{Cha07}.
\item When the distance is greater than a manually or automatically defined threshold (or conversely, the similarity is smaller than a threshold), characterize the corresponding snapshot as anomalous.
\end{itemize*}

When it comes to comparing consecutive graphs, there is no definite answer about the graph features that one should compare among the various timestamps. The novelty of each proposed algorithm lies in the ``graph summary'' it constructs, the distance/similarity function it uses, as well as the way it defines and chooses the threshold to flag an instance as anomaly. 
The majority of feature-extraction-based algorithms derive just a similarity score between two input graphs, without doing attribution; in other words, the algorithms usually cannot detect the nodes or regions of the graphs that changed most. 



\vspace{0.05in}
\noindent \emph{Approaches:}
\noindent \citep{ShoubridgeKWB02} and \citep{Bunke06} propose several ``graph footprints'' and metrics for monitoring  communication networks:
\begin{itemize*}
\renewcommand{\labelitemi}{$\bullet$}
\item Maximum Common Subgraph (MCS) distance of the adjacency 
or the ``2-hop'' matrices (=square of adjacency matrix),
\item error correcting graph matching distance \citep{ShoubridgeKR99}, which refers to the number of edit operations needed to convert a graph to another, and the costs of each operation may vary,
\item Graph Edit Distance (GED), which is a simplification of the previous distance, where only topological changes are allowed (i.e., no changes in edge weights),
\item Hamming distance for the adjacency matrices of the graphs, which essentially counts the number of different entries in the matrices,
\item variations of edge-weight distances,
\item $\lambda$-distance of the adjacency, the ``2-hop'', or Laplacian matrices, which is defined as the differences in the whole graph spectra, or the top-$k$ eigenvalues of the respective matrices. \citep{Peabody03} also proposes the $\lambda$-distance of the Normalized Laplacian matrices.
\end{itemize*}

 At this point, it is worth mentioning that although we consider $\lambda$-distance a graph-feature-based anomaly detection technique, it can be also classified as decomposition-based technique, since the extraction of the eigenvalues of a matrix is done by its decomposition (SVD \citep{GolubC96}, PCA \citep{Pearson1901}, LSI \citep{Deerwester90}, CUR \citep{DrineasKM06b}).

 \citep{ShoubridgeKWB02} and \citep{Bunke06} use the metrics for tracking sudden changes in communication networks for performance monitoring. The best approaches, in terms of change awareness, are the GED and MCS, both of which are NP-complete, but the former approach can be simplified given the application and it becomes linear on the number of nodes and edges in the graphs. In \citep{ShoubridgeKWB02}, the graph symmetric difference and difference in the vertex neighborhood subgraphs are proposed for change attribution.
 
 The authors in \citep{Bunke06} also go beyond the simple features, such as nodes, edges and weights, and introduce also more complex graph distance functions; the modality distance is defined as the Euclidean distance between the Perron vectors of the input graphs. Moreover, the authors propose the median graph distance; the median graph was first introduced by \citep{DickinsonBDK02}, and it is the graph that minimizes the sum of the edit distances to all the graphs in the sequence.

Two variations of GED with simple and non-linear cost functions for the allowed operations, which also accommodate the weights of the input graphs is given in \citep{KapsabelisDD07}, and used for accurate monitoring of dynamic computer networks. More details about the graph edit distance can be found in the survey \citep{GaoXTL10}.

In \cite{BunkeDIHK06}, the authors do not only compute the distances between consecutive graph instances, but all the pairwise distances (GED), and then apply an offline multidimensional scaling (MDS) procedure; each graph is represented by a point in the 2d-plane, and the distances between the points reflect their structural distances. This way the authors provide a nice, graphical representation of the changes that occur in a time-evolving graph; points that deviate from the mass of points correspond to anomalous timestamps or events.

\citep{GastonKW06} detect abnormal changes in time-evolving communication graphs using the diameter distance -- i.e., the difference in the graph diameter -- which is defined as the greatest of the longest shortest paths for all vertices. 


One of the early works in this category was conducted by \citep{Pincombe05}. The main idea of this work is to extract a single feature from each graph instance, and then, by using an appropriate metric, compare this feature in consecutive time ticks. Next, the resulting time series of the feature distances is modeled as an auto-regressive moving average process (ARMA) \citep{BoxJ90}, and the residuals (deviations from the model) are evaluated. The instances whose residuals exceed a threshold are considered anomalous. Briefly, ARMA is a model for describing time series by using two polynomials (the first for auto-regression, the second for moving average); it is widely used for predicting values in time series.
Among the 10 metrics that Pincombe used -- weight, maximum common subgraph (MCS) weight/edge/vertex, graph/median edit, modality, diameter, entropy, spectral distance --, the MCS edge, MCS vertex, edit, median and entropy were able to detect the anomalies that were introduced in a time-evolving IP traffic dataset.
Recently, another work that detects anomalies in time series (\textit{not graph data}), was introduced by Zhu and Sastry \citep{ZhuS11}. Their approach uses a General Likelihood Ratio (GLR) test based on Kalman filter for estimating the parameters of Auto-regressive Integrated Moving Average (ARIMA). The main insight remains the same; the detection of anomalies is based on the residuals of the filter, but in this case the monitoring of the residuals is done with the GLR test. Since this work is not used on graph data, we do not elaborate more here; however, it appears to be a nice alternative for the approach used in \citep{Pincombe05}.


Along the same lines, 
the authors in \citep{PapadimitriouGM08} introduce five graph similarity functions for directed, time-evolving web graphs: vertex/edge overlap similarity, vertex ranking, vertex/edge vector similarity, sequence similarity, and signature similarity. Among these metrics, the one that performs best in terms of change detection in web graphs is the Signature Similarity (SS), which is based on the SimHash algorithm. This algorithm uses as features the nodes and edges of the input graphs, weighted appropriately by their PageRank. 

\citep{BerlingerioKEF12} use a graph similarity approach for discontinuity detection in daily instances of social networks. 
In a nutshell, \textsc{NetSimile} consists of three phases: (i) Feature Extraction. The focus is on local and egonet-based features (e.g., number of neighbors, clustering coefficient, average of neighbors' degrees); (ii) Feature Aggregation. The node$\times$features matrix of the first phase is converted to a single ``signature'' vector that consists of the median, mean, standard deviation, skewness and kurtosis of each extracted feature over all the nodes in the graph; (iii) Comparison. The signature vectors are compared using the Canberra Distance, and a single similarity score is produced for consecutive timestamps of the graph sequence. The days that have low similarity score with the surrounding days are characterized as anomalous.

Another recent work, \citep{KoutraVF13}, proposes a complex graph-feature-based similarity approach, \textsc{DeltaCon}, for discontinuity detection, which enjoys several desired properties. The intuition behind the method is to compare the pairwise node affinities of consecutive snapshots of the graph sequence. These node affinities are computed in this work by a fast variant of Belief Propagation \citep{KoutraKKCPF11}. The matrices of pairwise node similarity matrices are then compared using the Matusita Distance (which is related to the Euclidean Distance), and the distance is finally transformed to similarity. A faster algorithm that avoids computing all the pairwise similarity scores is also proposed, and it is based on the idea of finding the similarity of all the graph nodes to non-overlapping groups of nodes (instead of each node individually). Once the time series of the consecutive-graph similarities is obtained, Quality Control with Individual Moving Range \citep{Montgomery97}
is used to spot the anomalous daily ENRON-graph instances.

In contrast to the most of the previous works that detect anomalous \textit{graph} instances, the following algorithms spot anomalous \textit{nodes} in a graph sequence.

The key idea in \citep{AkogluF08event} is the following: \emph{A node is anomalous at some time frame, if its ``behavior'' deviates from its past ``normal behavior''.} The authors build the ``behavior'' of the nodes by extracting various egonet node features (e.g., weighted and unweighted in- and out-degree, number of neighbors, number of triangles) from each snapshot of the graph sequence, and create a correlation matrix of node ``behaviors'' at each time window using Pearson's correlation coefficient. For each correlation matrix (one per time window), the principal eigenvector, which has one entry per node, is computed. 
 By placing all the corresponding entries of the eigenvectors in a vector, the ``eigen-behavior'' vector of each node is obtained, and compared against its typical ``eigen-behavior'', which is found by using averaging in the past time windows or SVD. The similarity between the ``behaviors'' is evaluated using the Euclidean dot-product. For low similarity between a node's ``behavior'' and its past ``behaviors'', the corresponding time window is reported as anomalous.

 Last but not least, the work of \citep{RossiGNH12} builds on top of \textsc{RolX} \citep{HendersonGETBAKFL12} --an NMF and MDL-based role extraction algorithm-- 
 to
 develop an algorithm that recursively extracts structural global and node features, and determine the nodes' roles (e.g., centers of stars, bridge nodes) over time. The authors use the method for understanding and tracking the network dynamics and evolution, but propose comparing the obtained node feature vectors over time in order to detect anomalous patterns. 
 Another similar approach, \textsc{DBMM} \citep{RossiGNH13}, that builds on top of \textsc{RolX} combines feature extraction, matrix decomposition, and a window-based analysis to model the node behavior in temporal graphs, predict future behaviors and spot anomalies. 
 First, the NMF and MDL-based role extraction algorithm computes the node group memberships. Then, by taking into account $k$ previous time steps, a role transition model per node is generated. 
 The approach does not detect anomalous graph instances, but anomalous {\emph nodes} per time step in decreasing order of anomalousness; the anomaly score of each node is defined as the difference between its estimated and true mixed membership.

%

\subsubsection{Decomposition-based Events}
\label{sec:eigenspaceBased}

\noindent \emph{Main idea:} The decomposition-based approaches  detect temporal anomalies by resorting to matrix or tensor decomposition of the time-evolving graphs, and interpreting appropriately selected eigenvectors, eigenvalues or singular values. The methods can be divided in two categories based on the representation of the graphs: matrices vs. tensors. 

\vspace{0.05in}
\noindent \emph{Approaches:}
We will first discuss the matrix-oriented approaches. These include the $\lambda$-distance \citep{Peabody03, ShoubridgeKWB02, Bunke06}, and the algorithms proposed in \citep{AkogluF08event} and \citep{RossiGNH13} that were presented in Section \ref{sec:dynamicfeatureextraction}. All of these approaches use graph features generated by SVD, eigenvalue decomposition or NMF, and, thus, can be also classified as decomposition-based anomaly detection techniques.

An additional work that handles each graph in the sequence separately by its matrix representation is 
\citep{Ide04} (also window-based approach), 
which aims at monitoring multi-tier Web-based systems. Conceptually, the method first extracts the principal eigenvector from the adjacency matrix of each graph; this is referred to as activity vector. Then, by applying SVD on the matrix that consists of the past activity vectors in a time window $w$, the typical activity vector is found, and the similarity between the current and typical activity vectors is computed as the cosine of the angle between them. The next step of the algorithm is to define the parameters of the von Mises-Fisher probability distribution \cite{fisher1993statistical} of the anomaly metric, and the threshold for characterizing a graph as anomalous or normal; the latter is found using an online algorithm. It is worth mentioning that the activity vector per node enables attribution, i.e. detection of the individual nodes that contributed most to the change in a particular graph instance. Based on \citep{Ide04}, the authors in \citep{IshibashiKHMKA10} detect uncommon traffic patterns in communication graphs. The novelty of their approach lies in the way the adjacency matrix of the network is created: 
instead of encoding the connectivity/communication patterns between the hosts, the cells hold the similarity between them, a property that is computed based on the overlap between their destination hosts.

SVD is not the only tool used by the decomposition-based detection algorithms. On the contrary, the last decade, several improvements on SVD have been proposed, including the CUR matrix approximation \citep{DrineasKM06b}, the Compact Matrix Decomposition (CMD) \citep{SunXZF08}, and Colibri-S \citep{TongPSYF08}. A pictorial comparison of the four methods is given in Fig.~\ref{fig:dynamic}. Given a set of 2-d data points, SVD constructs an optimal subspace using all the data points (full circles); CUR samples data points allowing for duplicates and linear redundancy (full circles), and approximates the original points based on them. 
CMD improves on CUR by sampling without substitution, while Colibri-S also guarantees that no linear redundancy exists in the sampled data points. Table \ref{tab:qualitative_comparison} provides a qualitative comparison of the four approaches. Although SVD is optimal in both norm-2 and Frobenius norm, it is inefficient time and space-wise. Moreover, the singular vectors do not have an intuitive interpretation since they describe the data in a rotated space, and the SVD of a matrix cannot be readily updated for dynamic or streaming graphs. CUR and CMD are much more efficient than SVD, and highly interpretable. Finally, Colibri-S is even more efficient in time and space, inherits the previous methods' interpretability, and additionally provides for efficient updates for dynamic graphs.

CMD \citep{SunXZF08} has been applied for anomaly detection in dynamic graphs:  the low-rank approximations of the sparse input graphs are used as their summaries. The reconstruction error of each graph from its summary 
is tracked over time, and, if it changes significantly at some time tick, the corresponding graph is deemed as anomalous.



\begin{figure}[h!]
\centering
   \includegraphics[width=0.95\textwidth]{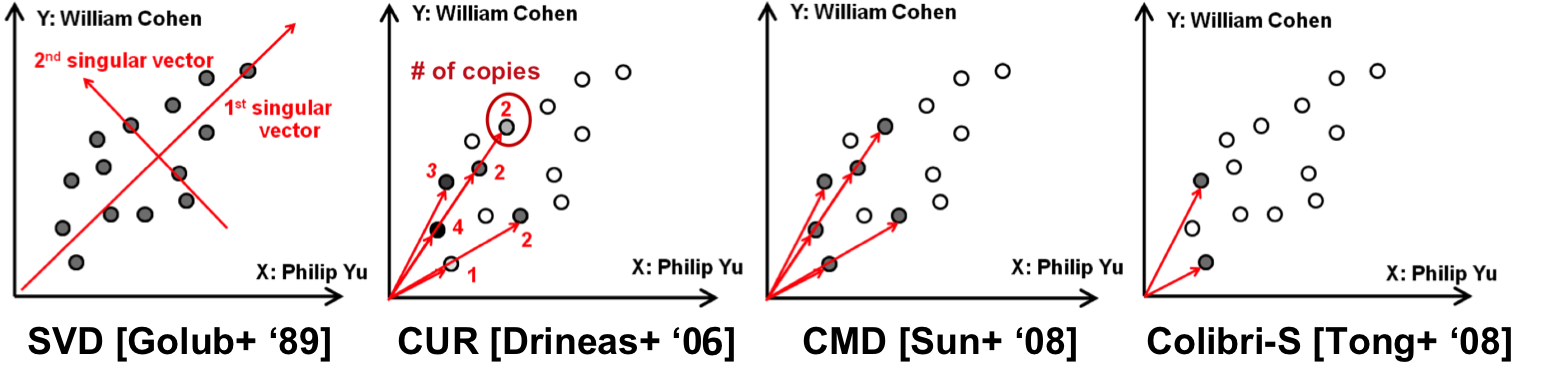}
\caption{Illustration of qualitative differences between matrix decompositions used for anomaly detection in dynamic graphs.}
\label{fig:dynamic}
\end{figure}

\begin{table}[h!]
\caption{Qualitative comparison of matrix decomposition methods: SVD, CUR, CMD, Colibri-S.}
\label{tab:qualitative_comparison}
\centering
{\footnotesize
\begin{tabular}{l||c|c|c} 
\toprule
 & SVD  & CUR/CMD & Colibri-S  \\ 
 \midrule
Quality & \cmark & \cmark & \cmark  \\
Efficiency  & \xmark & \cmark & \cmark \\
Interpretation & \xmark & \cmark & \cmark \\
Dynamic Graphs & \xmark & \xmark & \cmark \\ 
\bottomrule
\end{tabular}
}
\end{table}

Now we move on to the second category of decomposition-based event detection methods, which use tensors instead of matrices for the representation of the graphs.
Streaming Tensor Analysis (STA) \citep{SunTF06} is applied for anomaly detection to a computer network described by a source-destination-port graph. The authors introduce the tensor data structure, instead of a simple matrix, because they describe the networks with more entities than just source and destination.
Similarly to \citep{SunXZF08}, the main idea behind the proposed algorithm is to decompose the stream of tensors into projection matrices (one for each mode of the tensor), and incrementally update the latter matrices over time. If the incremental update leads at some point to high reconstruction error, then the tensor of that time stamp is considered anomalous. 

More recently, three more tensor-based approaches were proposed by \citep{KoutraPF12}, \citep{PapalexakisFS12}, and \citep{AraujoPGFBSPK14}. The first work simply uses the PARAFAC tensor decomposition; the second develops a fast, sampling-based, parallelizable decomposition algorithm for sparse tensors; the third, \textsc{Com2}, relies on tensor decomposition (PARAFAC) to obtain scores for time-evolving communities, and then applies MDL to find the ``important'' communities, and control their expansion (community size). In all three papers, for temporal anomaly detection, the first two dimensions of the tensors hold the information of the adjacency matrix, additional dimensions are used for attributes or extra entities, and the last dimension corresponds to the time. The detection of outlier groups of nodes at specific time stamps consists of observing different than 'usual' behavioral patterns in the factors of the decomposition (e.g. sudden increase in the interactions between nodes, bursty or bot-like behavior). 



%
%
%

\subsubsection{Community- or Clustering-based Events}
\label{sec:communityBased}

\noindent \emph{Main idea:} The main idea of the community or clustering-based approaches is, instead of monitoring the changes in the whole network, to monitor graph communities or clusters over time and report an event when there is structural or contextual change in any of them.

\vspace{0.05in}
\noindent \emph{Approaches:}
Being a building block for many applications, clustering, and the related, but not identical, problem of community detection, have been studied thoroughly in the data mining and theory communities: METIS \citep{KarypisK95}, one of the first partitioning algorithms that were developed, followed by its parallel implementation ParMETIS \citep{KarypisV96}; frequent subgraph mining \citep{KuramochiK2001}; spectral clustering \citep{ShiM97, NgJW01}; evolutionary clustering \citep{ChakrabartiKT06};  the Newman's algorithms for community detection in complex systems \citep{NewmanG03},\citep{New04c},\citep{Newman06062006}; co-clustering for concurrent clustering of the rows and the columns of the adjacency matrix of a graph \citep{DhillonMM03, Chakrabarti04}, and its distributed variants \citep{PapadimitriouSun08}; dynamic community detection algorithms \cite{Tantipathananandh07,conf/kdd/TantipathananandhB09,conf/icdm/TantipathananandhB11}, and empirical comparison of methods for  network community detection \citep{LeskovecLM2010}.


\textsc{GraphScope} \citep{SunFPY07} is an MDL-based, parameter-free algorithm for discovering node partitions in streaming, directed, bipartite graphs, and monitoring their evolution over time in order to detect events or changes. The partitions consist of ``similar'' nodes in the sense that splitting a partition leads to higher encoding cost of the adjacency matrix. The algorithm iteratively searches for the best source and destination partitions in each graph snapshot, until further partitioning does not lead to additional decrease of the encoding cost. Then, ``similar'' snapshots are merged into a segment and compressed together;  on the other hand, ``dissimilar'' consecutive snapshots lead to the creation of a new segment, and declaration of a change-point. A closely related tensor and MDL-based approach is \textsc{Com2} \citep{AraujoPGFBSPK14}, which tracks ``important'' communities over time, as described in Sec.~\ref{sec:eigenspaceBased}. Another approach that also uses node partitioning in order to identify structural anomalies in streaming graphs is \textsc{GOutlier} \citep{AggarwalZY11}, where the focus is on undirected, unipartite graphs. A reservoir sampling method is applied to create several node partitions and develop a structural edge generation model per partition, which describes the likelihood fit of an edge. Each edge in the incoming graph is characterized by its composite likelihood fit, which is defined as its median likelihood fit across all node partitions. Then, the graph's outlier score is represented by the geometric mean of all the composite edge likelihood fits, and the graph is reported as anomalous if its score is $t$ standard deviations below the average outlier score of the graphs seen so far.

A slightly different approach than the ones described above is the Bayesian anomaly detection method presented in \citep{HeardWPH10}. The authors focus on detecting anomalous regions in social networks using a two-stage Bayesian approach. At the first step of the method, the anomalousness of each edge is computed by modeling the interactions between each pair of nodes as a counting process. Also, at every graph instance, a p-value --based on the Bayesian learning of the count distributions-- is calculated for every existent edge and used in order to decide whether it is anomalous or not. The algorithm treats the graph sequence as a stream; it detects changes in the new graphs based on the history (sequential analysis), but also updates the history in light of the new instance (retrospective analysis). This step bears similarities with the methodology followed in \citep{AggarwalZY11}. However, the second and last step of the approach in \citep{HeardWPH10} is different; it essentially applies clustering techniques on the small subgraph consisting of the anomalous nodes and edges of the first step, so that locally anomaly regions are discovered.

A probabilistic modeling approach to change-point detection proposed in \citep{DBLP:journals/corr/PeelC14} uses the generalized hierarchical random graphs (GHRG) to model the community structure of real-world networks. The GHRC model decomposes
the nodes of the graph into a collection of nested groups, the 
relationships of which are represented by a dendogram. This representation captures the community structure at all scales. The change-points are identified by significant changes in the parameters of the fitted model through a generalized
likelihood ratio test.

Finally, \citep{GuptaGSH12} introduce the novel problem of detecting nodes which, over time, behave differently from the rest community members; those nodes are called \emph{evolutionary community outliers}. The approach, \textsc{ECOutlier}, consists of two parts: matching the time-evolving communities (which are detected in each graph instance by applying state-of-the-art techniques), and detecting the evolutionary community outliers. To solve the problem, an optimization framework that applies a coordinated descent algorithm is used to match the communities over time by appropriately weighting the contribution of the outlier nodes. It operates on pairs of consecutive timestamps of graphs, and returns a ranked list of community outliers.



\subsubsection{Window-based Events}

\noindent \emph{Main idea:} The last category of time-evolving graph anomaly detection algorithms encompasses methods that are bound to a time window in order to spot anomalous patterns and behaviors in the input graph sequence. Essentially, a number of previous instances are used to model the ``normal'' behavior, and the incoming graph is compared against those in order to characterize it as normal or anomalous.

\vspace{0.05in}
\noindent \emph{Approaches:}
 In \citep{PriebeCM05}, the authors apply scan statistics (as well known as ``moving window analysis'') to detect graph snapshots that have unusually high connectivity compared to the past. In general, scan statistics are used for detecting clusters of events in time and space \citep{Glaz07, Naus82, Kulldorff97}. Essentially, a local statistic is computed for each time window, and the maximum statistic within each window is called scan statistic; if the scan statistic exceeds a threshold, the corresponding time frame is deemed outlier.
 In this work, the locality statistic used on the disjoint, weekly snapshots of the ENRON who-emails-whom graph is the number of edges in the $k$-step neighborhood of each node, where $k=0,1,2$. This work is followed by a similar, scan-statistics-based approach in \citep{Neil11}, where model-based locality statistics are computed in paths and stars, instead of $k$-step neighborhoods. The method aims at spotting anomalies in computer networks, and the considered shapes are motivated by hacker attacks seen in real networks.



More recently, \citep{MongioviBRSPF13} tackled the problem of detecting contiguous regions in graphs that are anomalous over time by relating it to the NP-hard problem of finding the Heaviest Dynamic Subgraph (HDS). For each weighted graph in the input sequence, the anomalousness of each edge is computed as its statistical p-value using the empirical distribution of the edge weights; lower p-value corresponds to higher anomalousness. The proposed iterative algorithm, which solves approximately the HDS problem, alternates between the detection of the subgraph that maximizes the anomaly score for a given interval (spatial), and the detection of time interval that maximizes the score for a given subgraph (temporal). The output of the method is the regions that are more anomalous than a user-defined threshold. An interesting connection is observed between this work and \citep{HeardWPH10}; the approach in the latter paper can be used to compute the anomaly score of each edge, and then the algorithm in \citep{MongioviBRSPF13} can be applied to detect regions that demonstrate anomalous behaviors.

As mentioned in Sec.~\ref{sec:eigenspaceBased}, the method described in \citep{Ide04} can also be considered window-based, as the current activity of each node is compared against its activity in the past $w$ time ticks. Similarly, \citep{RossiGNH13} belongs to this category as well, since it models the role transitions of the nodes by taking into account the transitions from a number of previous time steps.
In addition, the probabilistic graph model fitting approach by \citep{DBLP:journals/corr/PeelC14} of Sec. \ref{sec:communityBased} is also a window-based one, where the generalized
likelihood ratio test is applied over a sliding window of fixed
length $w$ to detect if any changes have occurred with respect
to the fitted model.


\subsection{Discussion}

In the previous sections, we review the works in the literature that deal with the problem of graph anomaly detection over time. No matter which type of events are detected, the notion of graph or subgraph/community/cluster similarity usually comes into play at some step of the algorithms. Although the material that follows is not specifically designed for graph anomaly detection, it is closely related to it, as it gives alternative ways of computing the similarity between graphs, or, equivalently, their adjacency matrices.
\begin{itemize*}
\item{\textbf{Edit distance/graph isomorphism.} } One approach to graph comparison when the correspondence between the nodes in \textit{not} known is graph isomorphism. The underlying idea is that two graphs are similar if they are isomorphic \citep{graph_isomorphism}, or one is isomorphic to a subgraph of the other \citep{subgraph1} \citep{subgraph2}, or they have isomorphic subgraphs. The drawback of this approach is that the exact versions of the algorithms are exponential and, thus, not readily applicable to the continuously increasing in size and volume graphs. The graph edit distance \citep{edit_dist1}, which has been mentioned in Sec.~\ref{sec:dynamicfeatureextraction}, is a generalization of the graph isomorphism problem. 
\item{\textbf{Iterative methods.} } The assumption behind the iterative methods is that ``two nodes are similar if their neighborhoods are also similar''. In each iteration, the nodes exchange similarity scores and this process ends when convergence is achieved. Several successful algorithms belong to this category: the similarity flooding algorithm \citep{Melnik02} applies in database schema matching; 
SimRank \citep{SimRank02} measures the self-similarity of a graph, ie. it assesses the similarities between all pairs of nodes in one graph; 
the algorithm proposed by \citep{ZagerV08} introduces the idea of coupling the similarity scores of nodes and edges in order to compute the similarity between two graphs when the node correspondence is unknown. 
 \citep{MP11} develop two approximate sparse graph matching algorithms using message passing algorithms, and specifically Belief Propagation. Finally, \citep{KoutraTL13} design an alternating projected gradient descent algorithm for efficiently aligning big \textit{bipartite} graphs by exploiting the structural properties of the input graphs. 
\item{\textbf{Feature Extraction.} } A number of graph similarity functions, which have been used for graph clustering, classification and applications other than change-point detection, have been proposed in the literature. The research directions in this category include: algebraic connectivity \citep{Fiedler73} \citep{WilsonZ08}), a spectral method that has been studied thoroughly; an SVM-based approach on global feature vectors \citep{LiSYZ11}; social networks similarity \citep{MacindoeR10} which is based on graph features that are of value from the social viewpoint; computing edge curvatures under heat kernel embedding \citep{ElghawalbyH08}; comparison of the number of spanning trees \citep{Kelmans76}; fast random walk graph kernel for unlabeled \citep{KangTS12} or labeled graphs \citep{KashimaKA03}; graph kernels \citep{VishwanathanSKB10}, which are used for computing the similarity between graphs (not nodes). We should note that graph kernels cannot do attribution --i.e., detect the nodes that contribute most to a change in the graph sequence. 
\end{itemize*}

As in Section \ref{sec:static}, we close this section by comparing the dynamic-graph anomaly detection algorithms qualitatively, as well as quantitatively in Table \ref{tab:dynamic_qualitative}.

Choosing one of the algorithms presented in the previous sections for an anomaly detection application is not an easy task nor is there a unique appropriate algorithm; among the things that one should consider when choosing an algorithm are: the type of application (e.g., traffic, communication, computer network), the type of data at hand (e.g., weighted, unweighted, attributed), whether the correspondence between the nodes in consecutive graph snapshots is known or not, the time and parameter requirements, as well as the target of the application (detection of anomalous graph instance, subgraph, or node). Table~\ref{tab:dynamic_qualitative} can help refine the algorithms that can be applied in each case. The reader should bear in mind that, in many cases, applying multiple change-point detection techniques is meaningful, as it contributes to the discovery of different types of anomalies.

\begin{landscape}
\begin{table}[tbh]
\centering
\footnotesize
 \caption{Qualitative and quantitative comparison of anomaly detection algorithms for \textit{dynamic} graphs. The first four columns refer to the type of input graphs (with or w/o weights on the edges, with or w/o attributes (or labels) for the nodes);
 ``Linear'' holds true for those methods that have time complexity linear in the size of the input graphs (and false otherwise);
 ``Parameter-free'' methods correspond to those that do not expect any user-specified input parameters (+: parameter can be set, but is not required);
 ``Output format'' corresponds to the output type/format of the method (e.g. anomaly scores and their ranges);
 ``Node corresp.'' is true if the algorithm assumes that the correspondence between the nodes of the graph sequence is known;
 ``Attribution'' holds true if the algorithm spots nodes/edges/regions of graph that are anomalous (and false if it detects anomalous graph instances);
and ``Visualization'' refers to the graphical means used --if any-- to present the anomalous instances to the user (e.g., distribution plots, graph with the anomalous nodes/edges annotated). }
\label{tab:dynamic_qualitative}
\scriptsize{
\begin{tabular}{|l||c|c|c|c||c||c||c||c||c||c|} \hline
\multicolumn{1}{|g||}{\textbf{Algorithm}} &
\multicolumn{1}{|g}{{\tiny \begin{sideways}\textbf{Unweighted} \end{sideways}}} &
\multicolumn{1}{|g}{{\tiny \begin{sideways}\textbf{Weighted} \end{sideways}}} &
\multicolumn{1}{|g|}{{\tiny \begin{sideways}\textbf{Plain} \end{sideways}}} &
\multicolumn{1}{|g||}{{\tiny \begin{sideways}\textbf{Attributed} \end{sideways}}} &
\multicolumn{1}{|g||}{{\tiny \begin{sideways}\textbf{Linear} \end{sideways}}} &
\multicolumn{1}{|g||}{{\tiny \begin{sideways}\textbf{Parameter-free} \end{sideways}}} &
\multicolumn{1}{|g||}{{\tiny \textbf{Output Format}}} &
\multicolumn{1}{|g||}{{\tiny \begin{sideways}\textbf{Node corresp. } \end{sideways}}} &
\multicolumn{1}{|g||}{{\tiny \begin{sideways}\textbf{Attribution } \end{sideways}}} &
\multicolumn{1}{|g|}{{\tiny \textbf{Visualization (plot over time) }}} \\
 \hline \hline
\multicolumn{1}{|g||}{MCS {\tiny \citep{ShoubridgeKWB02, Bunke06}}}
& \cmark & \cmark & \cmark & \xmark &  \xmark & \cmark &  $[0,1]$  & \xmark &  \xmark   & consec. graph dist. scores \\  \hline
\multicolumn{1}{|g||}{HD {\tiny \citep{ShoubridgeKWB02, Bunke06}}}
& \cmark & \cmark & \cmark & \xmark & \cmark & \cmark &  $[0,1]$  & \xmark  & \xmark   & consec. graph dist. scores  \\  \hline
\multicolumn{1}{|g||}{ECGM {\tiny \citep{ShoubridgeKR99,ShoubridgeKWB02, Bunke06}}}
& \cmark & \cmark & \cmark & \xmark & \xmark & \cmark &  $[0,?)$  & \xmark  &  \xmark  & consec. graph dist. scores \\  \hline
\multicolumn{1}{|g||}{GED {\tiny \citep{ShoubridgeKWB02,Bunke06}}}
& \cmark & \xmark & \cmark & \xmark & \cmark & \cmark & {\tiny $[0,\#nodes+\#edges]$}  & \xmark  &  \cmark  & spy plot of graph difference   \\  \hline
\multicolumn{1}{|g||}{$\lambda$-distance {\tiny \citep{ShoubridgeKWB02,Bunke06}}}
& \cmark & \cmark & \cmark & \xmark & \xmark & \cmark+ &   $[0,\infty)$  & \cmark &  \xmark   & consec. graph dist. scores    \\  \hline
\multicolumn{1}{|g||}{GED\_w {\tiny \citep{KapsabelisDD07}}}
& \cmark & \cmark & \cmark & \xmark & \cmark &  \cmark  &  $[0,\infty)$  &  \cmark &   \xmark   & consec. ged scores \\  \hline
\multicolumn{1}{|g||}{Diameter Distance {\tiny \citep{GastonKW06}}}
& \cmark & \cmark & \cmark & \xmark & \xmark & \cmark & $[0,\infty)$  & \cmark &  \xmark  & consec. graph diameter distance \\  \hline
\multicolumn{1}{|g||}{MDS {\tiny \citep{BunkeDIHK06}}}
& \cmark & \xmark & \cmark & \xmark & \xmark & \xmark &  pairwise dist. & \cmark & \xmark  & MDS + consec. ged scores \\  \hline
\multicolumn{1}{|g||}{VEO {\tiny \citep{PapadimitriouGM08}}}
& \cmark & \cmark & \cmark & \xmark & \cmark & \cmark+ & $[0,1]$  & \cmark &  \xmark & consec. graph sim. scores \\  \hline
\multicolumn{1}{|g||}{Vertex Ranking {\tiny \citep{PapadimitriouGM08}}}
& \cmark & \cmark & \cmark & \xmark & \cmark & \cmark+ &  $[0,1]$  & \cmark &  \xmark  & consec. graph sim. scores \\  \hline
\multicolumn{1}{|g||}{Vertex/Edge Vector Sim. {\tiny \citep{PapadimitriouGM08}}}
& \cmark & \cmark & \cmark & \xmark & \cmark & \cmark+ &  $[0,1]$  & \cmark &  \xmark  & consec. graph sim. scores \\  \hline
\multicolumn{1}{|g||}{Sequence Sim. {\tiny \citep{PapadimitriouGM08}}}
& \cmark & \cmark & \cmark & \xmark & \cmark & \cmark+ &  $[0,1]$  & \cmark &  \xmark  & consec. graph sim. scores  \\  \hline
\multicolumn{1}{|g||}{Signature Sim. {\tiny \citep{PapadimitriouGM08}}}
& \cmark & \cmark & \cmark & \xmark & \cmark & \cmark+ &  $[0,1]$  & \cmark &  \xmark  & consec. graph sim. scores  \\  \hline
\multicolumn{1}{|g||}{{\tiny \citep{AkogluF08event}}}
& \xmark & \cmark & \cmark & \xmark & \xmark & \cmark & Z-scores & \cmark &  \cmark   & node Z-scores \\ \hline
\multicolumn{1}{|g||}{\textsc{NetSimile} {\tiny \citep{BerlingerioKEF12}}}
& \cmark & \xmark & \cmark & \xmark & \cmark & \cmark & $[0,1]$ & \xmark &  \xmark & consec. graph sim.scores \\  \hline
\multicolumn{1}{|g||}{\textsc{DeltaCon} {\tiny \citep{KoutraVF13}}}
& \cmark & \cmark & \cmark & \xmark & \cmark & \cmark+ & $[0,1]$ & \cmark & \xmark  & consec. graph sim. scores \\ \hline
\multicolumn{1}{|g||}{\textsc{Role-Dynamics} {\tiny \citep{RossiGNH12}}}
& \cmark & \cmark & \cmark & \cmark & \cmark & \cmark &  role memberships & \cmark &  \cmark  &  role memberships  \\ \hline
\multicolumn{1}{|g||}{\textsc{DBMM} {\tiny \citep{RossiGNH13}}}
& \cmark & \cmark & \cmark & \xmark & \cmark & \cmark &  role memberships & \cmark &  \cmark  &  role memberships  \\ \hline \hline
\multicolumn{1}{|g||}{\textsc{Eigen-space based} {\tiny \citep{Ide04}}}
& \cmark & \cmark & \cmark & \xmark & \xmark & \xmark & dissim. score $[0,1]$ & \cmark & \cmark   & sim. scores \& activ. vector change \\ \hline
\multicolumn{1}{|g||}{\textsc{STA} {\tiny \citep{SunTF06}}}
& \cmark & \cmark & \cmark & \cmark & \xmark & \cmark+ &  reconstruction err. & \cmark & \cmark &  reconstruction error  \\ \hline
\multicolumn{1}{|g||}{\textsc{CMD} {\tiny \citep{SunXZF08}}}
& \cmark & \cmark & \cmark & \xmark & \xmark & \cmark+ & reconstruction err. (SSE) & \cmark & \cmark &  reconstruction error  \\ \hline
\multicolumn{1}{|g||}{\textsc{ParCube} {\tiny \citep{PapalexakisFS12}}}
& \cmark & \cmark & \cmark & \cmark & \xmark & \cmark & factors & \cmark & \cmark & factors over time  \\ \hline \hline
\multicolumn{1}{|g||}{\textsc{GraphScope} {\tiny \citep{SunFPY07}}}
& \cmark & \xmark & \cmark & \xmark & \xmark & \cmark & reordered mat. spy plot & \cmark & \cmark  & encoding cost over time  \\ \hline
\multicolumn{1}{|g||}{\textsc{Com2} {\tiny \citep{AraujoPGFBSPK14}}}
& \cmark & \cmark & \cmark & \xmark & \cmark & \cmark & tensor decomp. & \cmark & \cmark  & tensor decomp. over time  \\ \hline
\multicolumn{1}{|g||}{\textsc{GOutlier} {\tiny \citep{AggarwalZY11}}}
& \cmark & \cmark & \cmark & \xmark & \cmark & \cmark & likelihood [0,1] & \cmark & \cmark  & likelihood over time  \\ \hline
\multicolumn{1}{|g||}{\textsc{Bayesian Approach} {\tiny \citep{HeardWPH10}}}
& \cmark & \cmark & \cmark & \cmark & \xmark & \xmark & p-values  & \cmark & \cmark  & predictive p-value  \\ \hline \hline
\multicolumn{1}{|g||}{\textsc{ECOutlier} {\tiny \citep{GuptaGSH12}}}
& \cmark & \cmark & \cmark & \xmark & \cmark & \xmark & community memberships  & \cmark & \cmark  & community memberships  \\ \hline \hline
\multicolumn{1}{|g||}{\textsc{Scan Statistics} {\tiny \citep{PriebeCM05}}}
& \cmark & \xmark & \cmark & \xmark & \cmark & \cmark &  scan statistics  & \cmark &  \cmark & scan stat. \& vertex scores  \\ \hline
\multicolumn{1}{|g||}{\textsc{Scan Statistics} {\tiny \citep{Neil11}}}
& \cmark & \cmark & \cmark & \xmark & \cmark & \cmark &  scores of regions  & \cmark &  \cmark    & scan stat. \\ \hline
\multicolumn{1}{|g||}{\textsc{NetSpot} {\tiny \citep{MongioviBRSPF13}}}
& \xmark & \cmark & \cmark & \xmark & \cmark & \xmark &  scores of regions & \cmark &  \cmark  & scores of regions  \\ \hline \hline
\end{tabular}
}
\end{table}
\end{landscape}

\section*{Concluding Remarks: Static \& Dynamic Graph Anomaly Detection}

\noindent \textbf{Evaluation.} To finalize Sections \ref{sec:static} and \ref{sec:dynamic}, we discuss evaluation methodologies of the graph-based anomaly detection approaches that have been employed in the literature thus far. Recall that ground truth data is often inexistent in the anomaly detection scenarios, thus, various methods in the literature have been evaluated in several different ways which we describe next. 

\vspace{-0.1in}
\begin{itemize}
\item {\em Internal evaluation.} This kind of evaluation mechanism uses the anomalousness scores of objects assigned by a given method to statistically quantify their extremity, e.g. by computing their $p$-values under the empirical distribution of scores of all objects.
This evaluation is internal, since the scores are dependent on the specific method and can be as diverse as likelihoods, compression costs, distances, etc., and may not necessarily directly tied with the external purpose of the anomaly detection.  

\item {\em Qualitative evaluation.}
Unlike the previous approach which is quantitative, qualitative evaluation employs informal procedures. One approach is to try to explain away the detected anomalies through a story related to a real-world scenario. Another approach is to incorporate domain knowledge to exploit and make sense of the detected anomalies. This latter methodology is often used in medicine, where the anomalies may help in knowledge discovery and help with diagnosis.

\item {\em Synthetic graph generation.}
A mechanism that is well resorted to is synthetic data generation.
In graph-based anomaly detection, several methods create realistic graphs using  graph generators such as preferential attachment \cite{barabasi99emergence}, Forest Fire \cite{LeskovecKF05}, random-typing graphs \cite{Akoglu2009pkdd} (power-law graphs), and the Waxman (Internet AS topology graphs) \cite{conf/mascots/MedinaLMB01} models.
Often, the kind of anomalies are directly injected to the synthetic graphs.
Sometimes the graph structure can also be  modified by randomly rewiring edges or swapping node attributes. The methods are then evaluated by their precision and recall in recovering the created anomalies. Synthetic graphs also help with evaluating the behavior of the proposed methods, such as their accuracy and scalability with changing graph characteristics, such as size and degree-distribution.

\item {\em Anomaly injection.}
The injection of synthetic anomalies has been discussed above. This is similar, only this time the anomalies are injected into the real-world graphs. One challenge in this version of anomaly injection is that the evaluation based on precision and recall becomes tricky, as it would be severe to call the anomalies detected other than the injected ones as false positives, given that the original graph may also contain same type of anomalies.

\item {\em Validation by external source.}
Another evaluation approach relies on multiple information sources that are consistent with each other in identifying the anomalies. In such a setting, one or more sources are used for the actual anomaly detection task. The detected anomalies are then tried to be validated or justified based on the rest of the unused information sources. For example, one may only use the graph structure to detect opinion spam and find out fake reviewers, and then use their temporal behavioral information, such as number of reviews written in a day, to see if the detected reviewers also exhibit suspicious behavior.
\end{itemize}

\noindent \textbf{Summary.} 
Finally to summarize, we create Table \ref{tab:summary23} including the various methods discussed this far under different categorization schemes
such as static and dynamic, and plain and attributed graphs.
Interestingly, we were unable to find any examples of methods that aims to find anomalies in dynamically changing attributed graphs.
We foresee that this would require novel definitions of anomalies in such a setting as well as necessitate the identification of real world scenarios
in which such definitions come alive.
Moreover, we notice that methods on static graphs strictly either deal with plain or attributed versions of graphs.
It would be interesting to build methods that can work with both; which apply to plain graphs but also can use side (attribute) information if available.
We classify those areas of research as open problems in our categorization,
and point them out as possible avenues for future exploration.

\begin{table}[!h]
\centering{
\caption{\normalsize Categorization of graph-based techniques discussed in Section \ref{sec:static} and \ref{sec:dynamic}.}
\label{tab:summary23}
{\scriptsize
\begin{tabular}{|l||c|c|} \hline
& {\normalsize \textbf{Plain}} & {\normalsize \textbf{Attributed}} \\ \hline \hline
\multirow{13}{*}{{\normalsize \textbf{Static}}} & & \\
&  {\small \textbf{[Section \ref{sec:staticplain}]}}  &  {\small \textbf{[Section \ref{sec:staticattributed}]}}  \\
& \textsc{Autopart} \citep{Chakrabarti04}  & \textsc{Subdue} \citep{NobleC03}\\
& \citep{conf/icdm/SunQCF05}   & \citep{conf/sdm/LiuYYHY05} \\
& \textsc{SCAN} \citep{SCAN1281280}  & \textsc{Subdue} \citep{conf/icdm/EberleH07}\\
& \textsc{OddBall} \citep{Akoglu2009oddball} &  \textsc{CODA} \citep{conf/kdd/GaoLFWSH10}\\
& \textsc{gSkeletonClu} \citep{conf/icdm/SunHHDZF10}  & \citep{conf/cikm/DavisLMR11} \\
& \textsc{NrMF} \citep{Tong@SDM11}  & \textsc{gOutRank} \citep{Muller13} \\
& \citep{conf/kdd/DingKBKC12}  &  \\ 
& \textsc{NetRay} \citep{KangLKF14}  &  \\ \cline{2-3}
& \multicolumn{2}{c|}{\multirow{2}{*}{{\normalsize{\textbf{Open}}}} }\\\\  
\hline \hline
\multirow{22}{*}{{\normalsize \textbf{Dynamic}}} & & \\
&  {\small \textbf{[Section \ref{sec:timeEvolvingGraphs}]}}  & {\small \textbf{[Section \ref{sec:timeEvolvingGraphs}]}} \\
& \citep{ShoubridgeKR99} & \textsc{STA} \citep{SunTF06} \\
& \citep{ShoubridgeKWB02} & \textsc{Bayesian App.} \citep{HeardWPH10}\\
& \citep{DickinsonBDK02} &  \textsc{Role-Dynamics} \citep{RossiGNH12} \\
& \textsc{Eigenspace-based} \citep{Ide04} & \textsc{TensorSplat} \citep{KoutraPF12}* \\
& \textsc{Scan Statistics}\citep{Pincombe05} & \textsc{ParCube} \citep{PapalexakisFS12}* \\
& \textsc{Scan Stat.} \citep{PriebeCM05} & \textsc{Com2} \citep{AraujoPGFBSPK14}* \\
& \citep{Bunke06} &  \\
& \textsc{MDS} \cite{BunkeDIHK06} &  \\
& \citep{GastonKW06} &  \\
& \textsc{GED\_w} \citep{KapsabelisDD07} &  \\
& \textsc{GraphScope} \citep{SunFPY07} & \normalsize{\textbf{Many Open}}\\
& \citep{PapadimitriouGM08} & \normalsize{\textbf{Challenges}}\\
& \citep{AkogluF08event} & \\
& \textsc{CMD} \citep{SunXZF08} & \\
& \citep{IshibashiKHMKA10} & \\
& \textsc{GOutlier} \citep{AggarwalZY11} & \\
& \textsc{ECOutlier} \citep{GuptaGSH12} & \\
& \textsc{NetSimile} \citep{BerlingerioKEF12} & \\
& \textsc{DeltaCon} \citep{KoutraVF13} & \\
& \textsc{NetSpot} \citep{MongioviBRSPF13} & \\ 
& \textsc{DBMM} \citep{RossiGNH13} & \\ \hline
\end{tabular}
}
}
\end{table}
\vspace{-0.2in}
{*: not applied in attributed graphs, but it is possible to admit labels/attributes.}

\clearpage

\section{Graph-based Anomaly Description: Interpretation and Sense-making}
\label{sec:description}

Like many other real applications, the ground-truth for graph anomaly either does not exist or is very difficult or costly to obtain. Consequently, the end analysts often have to spend much post-processing time to validate the detection results. For example, according to a recent DARPA BAA\footnote{\url{https://www.fbo.gov/utils/view?id=2f6289e99a0c04942bbd89ccf242fb4c}},  it is estimated that an intelligence agent can only perform 60 initial reviews on average for the so-called insider threat detection. This, coped with the facts that many graph anomaly detection algorithms (including insider threat detection) still have a high false positive rate, makes it extremely challenging and time consuming to identify at least one true positive in such applications. On the other hand,  it is usually much more persuasive for an ordinary user if the detection algorithm can tell not only which instance is abnormal, but also why it looks so different from the majority, normal examples.

To address these issues, graph anomaly attribution has been attracting more and more research attention in the recent years. In this section, we will review two main types of techniques. The first group aims to make the detection of each individual instance more `interpretable', which is usually done by encoding the so-called interpretation-friendly properties into the traditional graph anomaly detection algorithms. For this category, we will mainly use matrix-factorization based graph anomaly approach as an example. The second group tries to answer the following question. Given a set of initial suspects (e.g., the top ranked instances from a graph anomaly detection algorithm), how can we find and characterize the internal relationship among them so that we can better understand the root cause of such anomalies? For this category, we will introduce interactive graph querying and sense making.

\begin{definition}[Graph-based Anomaly Description Problem] \\
\indent \textbf{Given} a set of anomalies of graph entities (nodes and edges)  \\
\indent \textbf{Interpret and explain} the detection of the individual anomalies, \\
\indent \textbf{Find and characterize} the associations among the anomalies.
\end{definition}

\subsection{Interpretation-friendly Graph Anomaly Detection}


\noindent \emph{Main idea:} Here, we consider the first problem of how to make the detection of each {\em individual} instance (e.g., nodes, edges) more interpretable. The main idea is to {\em encode} the so-called interpretation friendly property into the traditional graph anomaly detection algorithms. We will present the matrix based graph anomaly detection methods. 

\vspace{0.05in}
\noindent \emph{Approaches:}
Suppose we have a bipartite graph (e.g., author-conference graph), and we can represent it by its adjacency matrix $\mat A$ with the rows being authors, columns being conferences and non-zero elements meaning the corresponding authors who have published papers in the corresponding conferences. In the matrix-based graph anomaly approaches, we start with factorizing the adjacency matrix as $\mat A = \mat{X} \mat{Y}'+ \mat{R}$. In this factorization, the two low-rank matrices $\mat X$ and $\mat Y$ usually capture the `normality' of the graphs (e.g., clusters, communities, etc); while the residual $\mat R$  measures the deviation from such `normality', and thus is often a good indicator of `anomaly'.

The different matrix-based graph anomaly detection approaches differ in the way they get these matrices. SVD/PCA is one of the most popular choices, where the columns of $\mat X$ and $\mat Y$ are the singular vectors (up to a scalar by the singular values) of the original matrix $\mat A$. While it is mathematically optimal in the sense that it minimizes the reconstruction error in both the L2 and the Frobenius norm, it is not necessarily good for interpretation.  We give two examples below to make such matrices less abstract and therefore more interpretable/consumable to the end analysts.

First, note that the singular vectors are usually the linear combination of {\em all} the columns/rows of the original adjacency matrix, which are not always easy for interpretation.
More recently, the so-called {\em example-based low-rank approximations} have started to appear,
 such as CX/CUR~\citep{DrineasKM06b}, CMD~\citep{Sun07@SDM} and Colibri~\citep{TongPSYF08}. All of these methods use the actual columns and rows
of the adjacency matrix $\mat A$ to form $\mat{X}$ and $\mat{Y}$.  The benefit is that they provide an intuitive as well as sparse representation, since
$\mat{X}$ and $\mat{Y}$ are directly sampled from the original adjacency matrix. The cost of such kind of decomposition is that
the approximation is often sub-optimal compared to SVD. We refer the readers to Section 3.2.2 for the detailed description of these methods. Also see Fig.~\ref{fig:dynamic} for a pictorial comparison. 

\hide{
\begin{figure}[htb]\label{fig:globalproj}
\begin{tabular}{cc}
\includegraphics[scale=0.35]{./illu_SVD.eps} & \includegraphics[scale=0.35]{./illu_CUR.eps}\\
(a) SVD & (b) CX/CUR\\
\includegraphics[scale=0.35]{./illu_CMD.eps} & \includegraphics[scale=0.35]{./illu_ColibriS.eps}\\
(c) CMD & (d) Colibri\\
\end{tabular}
\caption[An illustrative comparison between SVD, CX/CUR, CMD and Colibri.] {An illustrative comparison between SVD, CX/CUR, CMD and Colibri. Each dot is a data point in 2-d space. SVD (a) uses all the available data points to generate optimal projection directions. CX/CUR uses actual sampled data points (dark dots) as projection directions and there are a lot of duplicates. CMD (c) removes duplicate sampled data points. Colibri further removes all the linearly correlated data points from sampling.}\label{fig:globalproj}
\end{figure}
}


Another interpretation-friendly property that has been recognized widely in the recent years is {\em non-negativity} since negative values are usually hard to interpret. Non-negative matrix factorization (NMF) methods~\citep{Lee00@NIPS} which restrict the entries in $\mat{X}$ and $\mat{Y}$ to be non-negative have attracted a lot of research attention. By imposing such non-negativity constrains on the {\em factorized matrices}, NMF provides a more interpretable way for data mining tasks, e.g., clustering, community detection, etc. Note that although the NMF has been studied largely in the context of such applications (e.g., clustering), we would expect that it is also beneficial for graph anomaly detection, since it helps improve the interpretation of graph normality.

\begin{figure}[t]
\centering
   \includegraphics[width=0.9\textwidth]{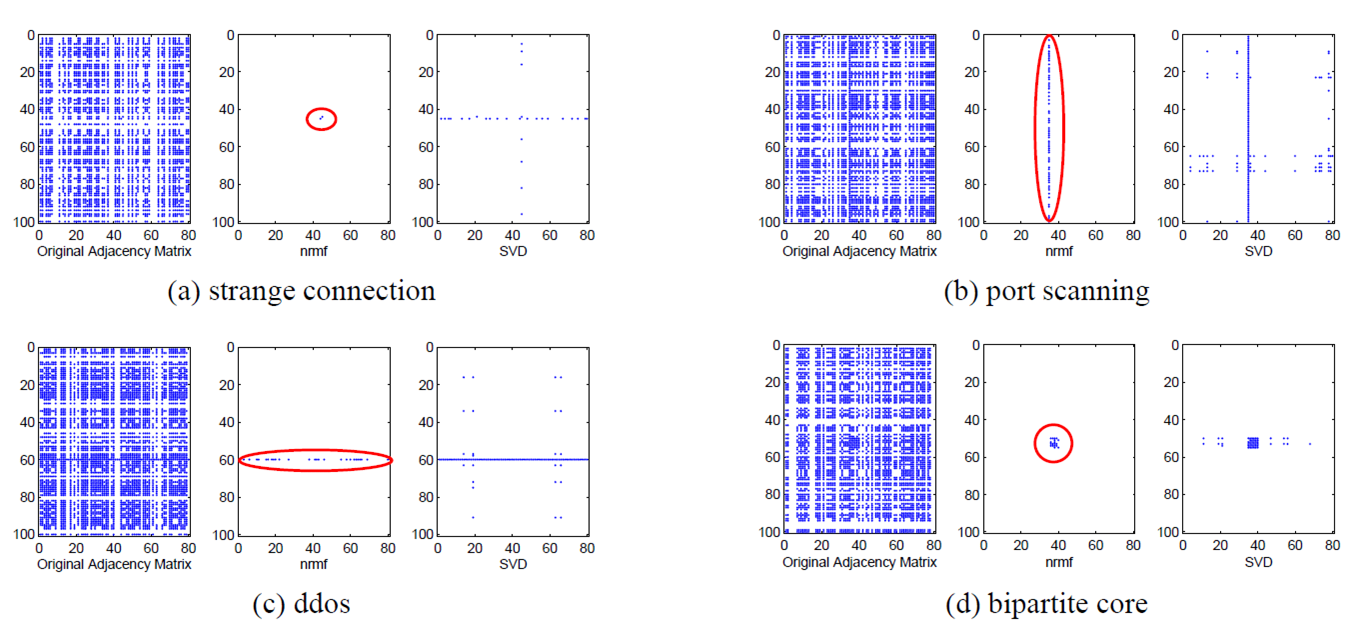}
\caption{Visual Comparison between NrMF and SVD. For each type of graph anomalies, the first column is the adjacency matrix of the original graph, the second and third columns are the residual matrices by NrMF and that by SVD, respectively.}
\label{fig:nrmf}
\end{figure}

In the context of graph anomalies, it is often the case that anomalies on graphs correspond to some actual behaviors/activities of certain nodes. For instance, we might flag an IP source as a suspicious port-scanner if it {\em sends packages} to a lot of destinations in an IP traffic network~\citep{Sun07@SDM}; an IP address might be under the DDoS (distributed denial-of-service) attack if it {\em receives packages} from many different sources~\citep{Sun07@SDM}; a person is flagged as `extremely multi-disciplinary' if s/he {\em publishes papers} in many remotely related fields in an author-conference network~\citep{Akoglu2009oddball};  in certain collusion-type of fraud in financial transaction network, a group of users always {\em give good ratings} to another group of users in order to artificially boost the reputation of the target group~\citep{ChauPF06}, etc. If we map such behaviors/activities (e.g., `sends/receives packages', `publishes papers', `gives good ratings', etc) to the language of matrix factorization, it also suggests that the corresponding entries in {\em the residual matrix} $\mat{R}$ should be non-negative. In order to capture such activities, non-negative residual matrix factorization (NrMF) has been proposed in~\citep{Tong@SDM11,Tong@SAM12}, which explicitly requires those elements in the residual matrix $\mat{R}$ to be non-negative if they correspond to actual links in the original graphs.
This in turn highly adds to the ease of interpretation. Fig.~\ref{fig:nrmf} presents a visual comparison between NrMF and the standard SVD on four typical graph anomalies~\citep{Tong@SDM11,Tong@SAM12}.

For feature-based graph anomaly detection, visualization in the (sub)space of the feature is also a very natural and powerful tool to improve the interpretation of graph anomalies \citep{Akoglu2009oddball,KangLKF14}. By making the abnormal graph nodes `standing out' in these low-dimensional plots, the end-user could have an intuitive understanding on which graph feature(s) makes them different from  normal.

\subsection{Finding the root cause of anomalies: Interactive Graph Querying}

\begin{figure}[t]
\centering
   \includegraphics[width=0.7\textwidth]{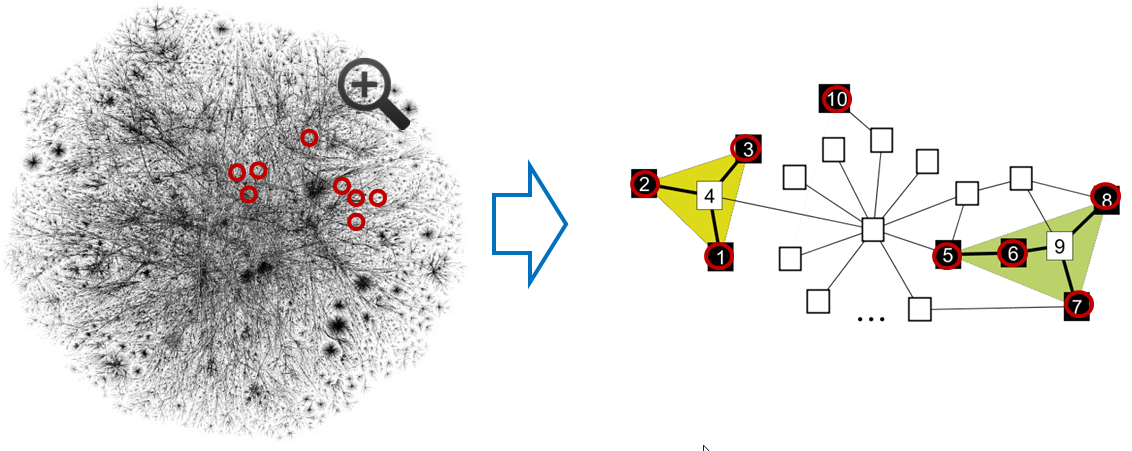}
\caption{The main idea of interactive graph querying. The left, given a set of detected abnormal nodes (red circles) from a given large graph (black). Right, the desired output which shows a concise summarization of these abnormal nodes (e.g., how they are further grouped into a few clusters, how the abnormal nodes within each group are linked to each other, etc).}
\label{fig:interactive}
\end{figure}

\noindent {\em Main idea:} Next, we consider the second problem of finding and characterizing the internal {\em relationships} among the anomalies so that we can better understand the root cause of such anomalies. We will introduce interactive graph querying. The main idea is to find a concise {\em context} where detected graph anomalies are linked to each other (See Fig.~\ref{fig:interactive} for an illustration). Note that while extremely useful in graph anomaly detection, these techniques themselves have a much broad applicability.

\vspace{0.05in}
\noindent \emph{Approaches:} {\em Connection subgraphs} is one of the earliest works along this line, which is defined as a small subgraph of a large graph that best captures the relationship from a source node to a target node~\citep{Faloutsos@KDD04}. The original method in ~\citep{Faloutsos@KDD04} is based on the so-called delivered
current. By interpreting the graph as an electric
network, applying $+1$ voltage to one query node and setting the
other query node $0$ voltage, it aims to choose the
subgraph which delivers maximum current between the query nodes. \citep{Koren@KDD06} propose using cycle-free effective conductance based method for this problem by only considering the top-k simple (i.e., cycle-free) paths from the source to the target. 
\citep{cartic05subgraph} further apply the
delivered current based method to multi-relational graphs.

Note that in all these works, they deal with pairwise source-target
queries. {\em Center-Piece Subgraphs} ({\sc CePS})~\citep{Tong@KDD06} generalizes this by considering the following settings: Given $Q$
query nodes in a social network (e.g., a set of top-ranked authors in a co-authorship network),
find the node(s) and the resulting subgraph, that have strong
connections to all or most of the $Q$ query nodes. This provides an intuitive tool to identify the potential root cause of graph anomaly detection results. For example, in the context of law enforcement, given a set of initial suspects, we may want to find other persons who have strong connections to all or most of the existing suspects, who might be the master criminal mind. The discovered path(s) in the resulting subgraph also provides an intuitive explanation on how/why the master mind connects to the individual suspects.

All the above works we have introduced in this subsection so far, assume, explicitly or implicitly, some specific connectivity structure among the query nodes. CePS provides certain degree of flexibility by allowing the so-called {\em k}-SoftAnd, where we only require the center-piece nodes to have strong connections to {\em k}-out-of-{\em Q} query nodes. But the end users still need to specify such a parameter {\em k} which is not necessarily an easy task for applications like graph anomaly detection. To address this issue, {\sc Dot2Dot}~\citep{Akoglu@SDM13,Chau@KDD12} proposes to find `right connections', that is, given a set of query nodes (e.g., the top-{\em k} ranked nodes in graph anomaly detection), it groups them into one or more groups and within each group, it finds the simple connections to characterize the relationship within that group. This problem itself is NP-Hard, and the authors propose efficient parameter-free algorithms to find approximate solutions. In the example of top-k ranking list from some graph anomaly detection algorithm, {\sc Dot2Dot} not only automatically groups the detected anomalies one or more groups and each group could correspond to a specific type of anomalies; but also provides some explanations why they belong to the same group and what is the possible root cause for that group of anomalies. Moreover, in the case there is a false positive node in the top-{\em k} ranking list (e.g., a node which is far away from all the other, true positive, nodes in the top-k ranking list) by automatically treating it as a group by itself.


\clearpage
\section{Graph-based Anomaly Detection in Real-world Applications}
\label{sec:fraud}

\begin{wrapfigure}[6]{r}{0.25\textwidth}
\vspace{-0.3in}
\includegraphics[width=80pt]{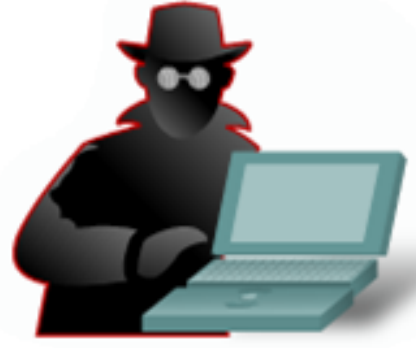}
\end{wrapfigure}
Next we shift our focus to real-world fraud and spam scenarios.
Several different techniques have been developed for fraud and spam detection in many real world scenarios including
frequent pattern mining \citep{conf/cikm/JindalLL10}, behavioral monitoring \citep{conf/kdd/FawcettP99}, supervised learning \citep{journals/sigkdd/PhuaAL04}, and so on.
In this section, we will motivate and focus on {\em graph-based} detection techniques for real-world applications and particularly highlight their advantages.
However, the purpose of our survey is not to suggest the superiority of graph-based techniques over other detection methodologies.
Rather, we introduce the available tools focusing on those that exploit graphs. It would be up to the application developers to carefully choose what tools suit their needs best as different approaches may achieve different performances depending on the application.
For a general survey on various fraud detection techniques, we refer the reader to \citep{Bolton2002Statistical}
and \citep{fraudcorr10}. 

We highlight two main advantages of graph-based fraud detection techniques as we discussed in Section \ref{intro};
relational nature of the problem domain and adversarial robustness. The former intuitively refers to the fact that fraud often occurs in two different ways,
(i) by word of mouth where the acquaintances of a fraudster can be considered as more likely to also commit fraud, and (ii) by collaboration where
closely related parties come together to commit fraud. In both scenarios, the relational ``closeness'' can be exploited with graph-based detection techniques.
The latter, robustness to adversaries, relates to the difficulty imposed on the attacker to break the detection method.
One can think that the graph-based representation of the domain in which fraud is committed is fully available only to the system administrators.
In other words, it is often the case that the fraudsters only have a limited view of the operational graph in which they act. Therefore,
it becomes harder for them to carefully cover their traces so as to ``fit in'' the global behavioral patterns of this graph.

In this part of the survey, 
we cover a wide range of applications
 including
 telecom fraud~\citep{journals/ida/CortesPV02},
auction fraud~\citep{PanditCWF07},
 accounting fraud~\citep{McGlohonBASF09},
 securities fraud~\citep{NevilleSJKPG05},
opinion spam~\citep{WangXLY11,conf/icdm/DaiZLP12},
trading fraud~\citep{conf/icdm/LiXLZ10}, 
network intrusion~\citep{Ide04,conf/kdd/DingKBKC12}, and
Web spam and malware detection~\citep{KangCF11,Gyongyi2004a,conf/mtw/WuGD06,conf/airweb/KrishnanR06,benczur05spamrank,becchetti2006link,conf/sigir/CastilloDGMS07}.

\subsection{Anomalies in telecommunication networks}
While there are many types of telecommunications fraud, one of the most prevalent is known as the subscription fraud.
In this type of fraud, the fraudster often acquires an account using false identity with the intention of using the service for free and not making any payments.

One of the earliest studies that proves the graph-based methods effective in telecommunications fraud detection is done by \citep{journals/ida/CortesPV02}, who mainly
use linkage analysis together with temporal and calling volume information.
In particular,
they build and maintain subgraphs around each phone account which they name as the ``communities of interest'' (COI) of the account.
The COI mainly contains the other phone accounts that are most related to the given account in terms of dynamically weighted measures that consider the call quantity and durations between these parties over time. Using these informative subgraphs updated daily,  two discriminative properties are observed. Firstly, fraudulent phone accounts are found to be linked; fraudsters either directly call each other or they call the same phone numbers which puts them in close proximity in the COIs. A second observation shows that it is possible to spot new fraudulent accounts by the similarity of their COIs to previously flagged fraudulent COIs---this is due to detected and disconnected fraudsters by the phone operator creating new accounts and exhibiting similar calling habits, which are effectively captured by their COIs.

These graph-based linking methods provide powerful machinery on top of previously used signature-based methods \citep{journals/datamine/CortesP01,conf/kdd/CortesFPR00},
where few simple measures such as extensive late night activity and  long call durations have been taken as indicators for fraudulent behavior.

 \subsection{Anomalies in auction networks}

 Auction sites such as eBay, uBid, bidz, and Yahoo! Shopping are attractive targets for auction fraud,
  which constituted about 25\% of the complaints to Federal Internet Crime Complaint Center (IC3) in the U.S. in 2008 \citep{IC32008}.
 The majority of online auction fraud occurs  as non-delivery fraud ($\sim$33\%), where the seller fails to deliver/ship the purchased goods to the buyer.

\citep{ChauPF06} developed one of the very first graph-based methods to spot fraudsters committing auction fraud and showed the effectiveness of their method on a large crawl of eBay data. The motivation to use graph-based methods in that domain is the insufficient solutions based on the individual's features, such as age, geo-location, login times, session history, etc. which are ``easy'' to fake.
As we discussed earlier in this section, as well as in Section \ref{sec:why}, the intuition is that as the fraudsters have only a local view of the auction graph, it is ``harder'' for them to alter their behavior and still be able to ``fit in'' this graph at large without knowing all the patterns of interactions.

The analysis of the fraudsters' behavior reveals that in order to game the feedback and reputation system, fraudster create additional accounts or ``roles'' called accomplices. Thus, fraudsters exhibit two roles:

\vspace{-0.05in}
\begin{itemize}
\item {\em accomplice}: trades with honest users, looks legitimate
\item {\em fraudster}: trades with accomplices to ``sell'' (cheap) items and receive good feedback to boost reputation, and occasionally commits fraud with honest users when reputation is high enough to convince them
\end{itemize}
\vspace{-0.05in}

Accomplices and fraudster do not necessarily interact among each other. Moreover, honest users trade among themselves as well as with accomplices that also look like honest users. As such, there is quite a bit heterophily among the labels of neighboring nodes: with fraudsters mostly linked to accomplices and occasionally to honest users, accomplices linked to both fraudsters and honest users serving as middle-men, and honest users mostly linked to other honest users and accomplices.

Using the insights of these interaction characteristics, \citep{PanditCWF07} developed a relational classification model based on RMNs that can capture these complex correlations (in particular heterophily) among the node labels (honest, accomplice, fraudster), and used LBP for inference.

\subsection{Anomalies in accounting networks}

Accounting fraud involves the task of spotting high-risk accounts with suspicious transactions behavior.
Many existing techniques for detection rely on (noisy) domain knowledge and rule-based signals, for example, based on large number of returns, many late postings, round-dollar entries, etc.

Based on the insight that closely related accounts by their transaction relations would be more likely to have the same labels (risky vs. non-risky), \citep{McGlohonBASF09} use relational classification to detect accounting fraud. Here, unlike the heterophily observation in \citep{ChauPF06}, the homophily (auto-correlation) of neighboring class labels is assumed. Similarly, a RMN representation is developed and LBP is used for inference.

One of the representational powers of global joint models like RMNs, in addition to their ability to capture complex correlations, is the fact that they can integrate prior knowledge if available. In this particular application, the prior knowledge (probability) of accounts being risky translates to prior belief potentials in the RMN representation. In fact, \citep{McGlohonBASF09} use the previously used (noisy) domain knowledge based on rule-based flags to estimate the prior beliefs. These beliefs are then propagated in the network where some of them are corroborated and some may be discarded.
Their results showed that through this type of graph-based validation, the detection (true positive) rate improved significantly over the rule-based methods for the same (small) false positive rate.

 \subsection{Anomalies in security networks}
 Relational learning has also been used in securities fraud detection where the task is to spot securities brokers that are likely to commit fraud and other violations  of securities regulations in the future.
 While previous methods used handcrafted rules based on information intrinsic to the brokers such as the number and type
 of past violations,  \citep{NevilleSJKPG05} exploited relational information such as social, professional,
 and organizational relationships (e.g. past co-worker) among the brokers.
 In fact, this is one of the applications where the likelihood of committing fraud is highly dependent on social phenomena: communicated and encouraged by word-of-mouth by people who wish to commit fraud that relational methods are excellent at spotting.

 In particular, \citep{NevilleSJKPG05} use a subgraph representation for each of the securities brokers.
 Each subgraph includes, in addition to the target broker,
 various types of other objects (e.g., firms, disclosures), as well as links that
 represent relationships between these objects (e.g., employment links between a broker and a branch, filing links of disclosures on the broker), and attributes
 on these objects and links.
They then
 learn relational probability trees \citep{Neville03Learning} which exploits (aggregated) relational features of those subgraphs to model the distribution of the class labels, showing that the learned models rank brokers in a
manner consistent with the subjective ratings of experienced examiners, and better than handcrafted rules.

 \subsection{Anomalies in opinion networks: deception and fake reviews}

 Review sites such as Yelp, TripAdvisor, Amazon, etc., are attractive targets for opinion spam.
 Opinion spam exhibits itself as hype or defame spam, where (often paid) fraud reviewers write fake reviews to untruthfully boost or damage a vendor's reputation, respectively and cause unjust perception of the services by future customers.

 This problem has been approached by three different methodologies, based on (i) behavioral analysis \citep{JindalL08,conf/cikm/JindalLL10,Feng12,XieWLY12}, (ii) language stylometry analysis to spot deception \citep{FengACL12,Ott12}, and (iii) relational analysis and network effects to exploit connections among fraudulent reviewers \citep{WangXLY11,WangXLY12,Akoglu2013opinionspam}.
More specifically, with respect to (i) and (ii), \citep{JindalL08,conf/cikm/JindalLL10} extract behavioral features such as review length, posting times, time order of reviews (whether first posted review or not), etc. in addition to rule-based mining to spot suspicious reviewers. \citep{Feng12} study the distributional patterns in rating behaviors, while \citep{XieWLY12} focus on temporal reviewing behaviors to detect fake review(er)s.
As for language-based detection, \citep{Ott12} unearth the excessive usage of superlatives, self-referencing, rate of misspell, and agreement words in fake reviews as important clues.

With respect to graph-based detection (iii), \citep{WangXLY11} developed a propagation algorithm to capture the relationships between reviewers, reviews, and stores (or products, services).
The method defines a trustiness score for each reviewer, reliability score for each store, and a honesty score for each review.
These scores are defined in terms of one another: reviewer trustiness is a (non-linear) function of his/her reviews' honesty scores,
store reliability is a function of the trustiness of the reviewers writing reviews for it, and finally review honesty is a function of the reliability of the store it is written for as well as the trustiness of the reviewers who have also written reviews for the same store it was written for.
The algorithm randomly initiates these scores, and updates them iteratively until some convergence criterion is reached.
This is similar in design to the HITS algorithm by \citep{Kleinberg98}  where the authoritativeness and hubness scores of Web pages, which are defined in terms of linear functions of each other, are updated iteratively. On the other hand, the algorithm is not guaranteed to converge, and cannot exploit extra knowledge such as textual clues or behavioral information but is complementary to these previous methods.

Most recently, \citep{Akoglu2013opinionspam} exploited relational classification for opinion spam detection.
In particular, they developed a relational model based on RMNs that can capture the correlations
between reviewers and stores, and used LBP for inference.
One main difference from earlier network classification based methods is the signed nature of the opinion network,
in which the reviewers are connected to stores (or products) with positive ($+$) or negative ($-$) links that capture the sentiment of their reviews (e.g., like/dislike).
The signed links affect the label correlations: e.g., while a fraudulent reviewer is likely to link to a low-quality store with a $-$ link (unjustly boosting its reputation),
it is less likely for him/her to link to a high-quality store with a $+$ link; although this latter case occurs where fraudulent users occasionally write truthful reviews to camouflage their
otherwise fraudulent activities, which is accounted for in the RMN model.



 \subsection{Anomalies in financial trading networks}
\citep{conf/icdm/LiXLZ10} use graph-based substructures and their efficient detection to spot potential fraudulent cases in trading networks.
These cases consist of a group of traders that trade among each other in certain ways so as to manipulate the stock market.
More specifically, the group of traders may perform transactions on a specific stock among
themselves for some amount of time during which the overall shares of the target stock in their trading accounts
 increase and they end up producing a large volume of transactions on this stock.
After the stock price goes up, these traders start selling the acquired shares to the public producing excessive volume of transactions to traders other than themselves.

These two different behaviors of a group of traders within consecutive time windows are formulated in graph-based terms.
In the former, in which excessive buying of the stock occurs, the in-link weights are expected to be quite high (these are called blackhole patterns), while in the latter selling stage the out-link weights highly exceed in-links' (these are called volcano patterns).
These two fraudulent trading behaviors are formally defined and formulated in graph-theoretic terms, and efficient algorithms are developed to detect such patterns quickly in very large and dynamically changing financial trading networks.



\subsection{Anomalies in the Web network: spam and malware}

One suitable way to define Web spam is any attempt to
get an unjustifiably favorable relevance or importance score
for some Web page, considering the page's true value.
One of the main techniques in combating spam and malware on the Web has been to use trust and distrust propagation
over the graph structure. These techniques assume that a link between two pages on
the Web signifies trust between them; i.e., a link from page
$i$ to page $j$ is a conveyance of trust from page $i$ to page $j$.
Moreover,
if the target page is known to be a
spam page, then they consider the trust judgment of the source
page as invalid, in which case the source page is penalized
for trusting an untrustworthy page.

One of the earliest methods in improving the PageRank algorithm to combat Web spam is
TrustRank~\citep{Gyongyi2004a}, which employs the idea of propagating trust from
a set of highly trusted seed sites.
Initially human experts select a list of
seed sites that are well-known and trustworthy on the Web.
Each of these seed sites is assigned an initial trust score. A
biased PageRank is then used to propagate
these trust scores to the descendants of these sites.
The amount of trust decreases with distance from the seed set and the
number of outgoing links from a given site.

Anti-TrustRank~\citep{conf/airweb/KrishnanR06} can be thought of as the dual of
TrustRank that performs propagation starting from known bad pages and propagate distrust instead of trust.
The intuition used in this work is that the pages
pointing to spam pages are very likely to be spam pages
themselves. Anti-Trust is propagated in the reverse direction along incoming
links, starting from a seed set of spam pages.

\citep{conf/mtw/WuGD06} also point out several issues regarding TrustRank's assumptions, such as the fact that
it looks at outgoing links  and divides trust propagated to children by their count, which causes two equally trusted pages (but with different number of children) propagate different trust scores to
their children. Moreover the children accumulate trust by simply summing the trust scores from their parents. Instead, they
use different splitting and accumulation techniques. In addition, they employ both trust and distrust propagation, and finally assign a weighted score of the two.

One of the main challenges of these methods discussed so far is that they all expect a manually labeled seed set of good or bad pages.
~\citep{benczur05spamrank} propose a novel way to overcome this challenge and fully automate the process.
Their idea is to look at the distribution of PageRank
scores of neighbors for each node (i.e. Web page in the graph), which is expected to be power-law distributed
given the overall PageRank score distribution being power-law and the self-similarity of the Web.
For those nodes where the PageRank distribution of their neighbors deviate significantly from power-law, they assign a ``penalty''.
Similar to Anti-TrustRank, a new PageRank biased by the penalty scores gives the Spam-Rank scores.

Link-based spam detection~\citep{becchetti2006link} looks at
different graph-based measures which are then used as features to train classification models.
The graph features include PageRank, TrustRank scores, degree, assortativity (i.e. degree
correlation), fraction of reciprocal edges, average degree of neighbors, etc.  These type of link-based features are
complementary to other techniques that use content-based features, such as the number of words, number of hyperlinks, text redundancy, etc. \citep{ntoulas06detecting,conf/www/CanaliCVK11},
and content-free features, such as URL-based-only host and lexical features \citep{conf/kdd/MaSSV09}.

The work called `Know your neighbors'~\citep{conf/sigir/CastilloDGMS07} makes use of various types of features in tandem to learn classifiers and furthermore, use the graph structure to ``smooth'' the classification results.  Main idea is to extract features that are (a) link-based
(edge-reciprocity, assortativity, TrustRank score, ratio of TrustRank to
Pagerank, radius, neighborhood growth rate for increasing number of hops, etc.); and
(b) content-based (compression ratio, entropy of n-grams, etc.).  Using these
features they learn classifiers (decision-trees) and then smooth the classifier scores using
the graph structure. In particular, they use three different ways to exploit the Web graph: (i) clustering where all the nodes in the same cluster is relabeled by the majority of the initial labeling,
(ii) random-walk-with-restart where probabilities are set to normalized spamicity scores from the classifier (similar to Anti-TrustRank), and (iii) stacking where a set of extra features
for each object are added to the classification over iterations by combining the predictions for the related
objects in the graph (this indeed is an ensemble method).

\subsection{Anomalies in social networks}

Related to the previous section, another group of malware detection methods focuses on social malware in social networks such as Facebook. Such malware is also called socware.
Socware consists of any posts appearing in one's news feed in social media platforms such as Facebook and Twitter
that (i) lead the user to malicious sites that compromise the user's device, (ii) promise false rewards and make the user perform certain tasks (e.g. filling out surveys) potentially for someone else's benefit, (iii) make the user boost the reputation of certain pages by clicking or `liking' them, (iv) make the user redistribute (e.g., by sharing/re-posting), and so on.

To combat socware, \citep{Rahman2012} propose a classification framework that exploits ``social-context-aware'' features,
such as message similarity of posts across different users who shared (or made to share as in (iv) above) a particular post, the size of the propagation of the post in the network,
the total `like' and comment counts of other network users on the post, etc. in addition to other content-based features.
In another study, \citep{gaofiltering12} perform online spam filtering on social networks using incremental
clustering, based on features that also include network-level features such as sender's degree and the interaction history between users.
These methods rely on learning classifiers based on collective feature sets (including graph-based features). On the other hand,
it would be interesting to see if unsupervised methods that directly focus on graph mining could help in identifying online socware, by
studying the propagation-based dissemination of socware in the network.

\subsection{Anomalies in computer networks: attacks and intrusion}

Most graph-based network intrusion detection methods focus on the dynamically growing and changing nature of the
network graph. In this graph, the nodes represent the agents in the networks, such as ad/file/directory servers and client nodes, and edges represent their communications over the network (note that the edges may be weighted, capturing volume or frequency). The insight behind tracking the dynamic nature of the network graph is the assumption that the communication behavior of a compute node would change when under attack.

There exist two main challenges associated with tracking large communications networks and the necessity to consider their relational characteristics:
(i) large number of compute nodes makes it impractical to monitor them individually, moreover the behavior of the nodes
may be dependent on each other and thus monitoring them in isolation would bypass their correlations; (ii)
large number of edges makes it impractical to study the highly dynamic time-series of communications volume in tandem.

For these reasons, \citep{Ide04} monitor what is called the ``activity'' vector of nodes. The activity score of a node is computed collectively; if a node links to many active nodes, its activity score is high. With this definition, the activity vector essentially becomes the principal eigenvector of the adjacency matrix that depicts the communication graph. They track this vector over time by measuring the change in its direction and magnitude and develop online thresholding techniques to decide when to flag a change as a significant event. These events may correspond to network attacks as well as failures and other network configuration changes.

\citep{SunXZF08} exploit matrix decomposition to capture the norm of network activity.
They employ a sparse and efficient (both in time and storage) method called Compact Matrix Decomposition to decompose the adjacency matrix of the network graph and use relative sum-square-error of reconstruction as a measure of change to track over time for newcoming snapshots of the network graph. They observe that this new measure of change detects events that total volume monitoring misses.

Another graph-based method \citep{conf/kdd/DingKBKC12} considers analysis of network communities,
as we discussed in Section \ref{sec:staticplain}. Simply reput, the idea is to monitor cross-community communication behavior to spot network intrusion. Intuitively, communications that cross community boundaries, considered as anti-social, are suspicious and can be treated as signal of attack. The ROC curves show that methods based on this insight achieve over 90\% accuracy in detection, however with a somewhat high false alarm rate of about 50\% in ground-truth data with malicious attacks.

 Finally, while not directly focusing on network intrusion, \citep{DBLP:conf/imc/2007,IliofotouKFMPV11} use graph based network traffic representations, called traffic dispersion graphs, to analyze, monitor, visualize, and classify network traffic.



\section{Conclusions and Open Challenges}
\label{sec:open}

\noindent
\textbf{Summary.}
In this survey, our aim has been to provide a comprehensive overview of graph-based techniques for anomaly, event, and fraud detection, as well as their use for post-analysis and sense-making in explaining the detected abnormalities.
Following our taxonomy in Figure \ref{fig:survey}, we surveyed quantitative detection and qualitative explanation/attribution techniques as two main parts. The detection methods are further categorized into three groups: (i) anomaly detection in static graphs; (ii) event detection in dynamic graphs; and (iii) fraud detection in real-world scenarios. The first two groups (anomalies and events) consist of {\em general} abnormality definitions and their detection techniques proposed mainly by the data mining community. 
The third group (fraud scenarios) consists of {\em specialized} techniques for particular fraud types as observed in the real world and mostly involve (machine) learning approaches.
Furthermore, the attribution techniques highlight graph-based tools for analysis, visualization, monitoring, exploration, and sense-making of the anomalies.

\vspace{0.075in}
\noindent
\textbf{Conclusions.}
One of the main messages we aimed to convey has been the expressiveness of graphs in capturing real world phenomena, which makes them a very powerful machinery for abnormality detection.  In particular, we emphasized that (i) data instances are often inter-dependent and exhibit long-range correlations, (ii)  the anomaly detection problem is often relational in nature (e.g., opportunistic or organized fraud), and (iii) robust, hard-to-circumvent machinery is essential in the arms race with the attackers in fraud scenarios.
As such, graphs prove to be effective in all these aspects.

Our aim, however, is not to claim the superiority of graph-based methods over other detection techniques. On the other hand, our goal is to highlight the advantages of graphs, and provide a comprehensive list of available algorithms and tools that exploit graphs to build anomaly detection solutions.
We believe that those would prove complementary to other types of techniques and should most probably be used in tandem for better detection performance. In fact, it is at the discretion of the practitioners to decide what type of scenarios best describe the problems they are dealing with, as well as what tools best fit their needs.

\vspace{0.075in}
\noindent
\textbf{Open Challenges.}
While there has been tremendous amount of work in developing graph-based algorithms and machinery for graph-based abnormality detection and attribution, we believe there is still more work that needs to be done. In this part, we provide a discussion of open challenges which we group in two parts: \textit{theoretical} and \textit{practical} research challenges.

\paragraph{Theoretical research challenges.}
While there has been considerable amount of work on static graphs, there still remain problems in the study of dynamic graphs.
\begin{itemize}

\item {\em Anomaly Detection on Attributed Dynamic Graphs.} While static attributed graphs have been exploited in abnormality detection, there exists only a few works on spotting anomalies by exploiting dynamic {\em attributed} graphs (see Table \ref{tab:summary23}). It is certainly of interest to develop definitions and formulations for abnormalities in such settings, and explore and identify where they could find applications in the real world.
\item {\em The History/Trace of Dynamic Updates.} While most techniques for dynamic graphs consider and work with edge/node updates,
there exists no work that exploits the {\em history} of the updates.
For example, imagine a Web page having a link to a malicious Web page in the past which is later removed.
While from the change point of view this is an edge removal, the existence of such a link in the history of the page should be taken into account in making future evaluations, rather than treating and committing to the change as a simple edge removal.
\item {\em Choosing the `Right' Time Window/Granularity.} Many algorithms for time-evolving graphs require a time-window for feature extraction or computation of the normal graph/node activity; one of the open questions is how to choose this window in order to discover the different types of outliers in the graph sequences. Would it make sense to set it to a day, or a week, or a month, based on the respective periodicities that have been reported for human activities/botnet attacks etc.? Would another time granularity serve for detecting the existent anomalies? Or, would a combination of time granularities work best?
\end{itemize}

Moreover, while there has been considerable work from {\em algorithmic} point of view in abnormality detection, there still remains problems from {\em systems} perspective.
\begin{itemize}
\item {\em Adversarial Robustness.} Most methods in the data mining and machine learning community focus on detection performance while ignoring {\em adversarial robustness}. It is of high interest, from the practitioner's point of view, to understand the adversarial robustness of a new algorithm; i.e. how easy is it to break the algorithm, or what is the minimum amount of knowledge or computational power the attacker needs to have access to, in order to camouflage his/her bad activities.
\item {\em The Cost of Graph Anomaly Detection.} Most methods ignore the {\em cost} aspects of information. These costs, on the other hand, may exhibit themselves in various forms with varying levels, e.g., cost in measurement and monitoring exerted on the system; cost in being exposed to certain types of attacks exerted on the users; and cost in getting around of the algorithms exerted on the adversaries (which also relates to the above). These varying costs should be accounted for differently in algorithm development.
\item {\em Scalable Real-time Discontinuity Detection.} One of the most important future challenges is to develop scalable approaches for real-time discontinuity detection, i.e., for streaming graphs. Specifically, research should focus on algorithms that are linear, or, even better, sub-linear to the input.
\end{itemize}

\paragraph{Practical research challenges.}

Challenges from the practitioner's point of view, which could also be posed as research problems, include the following.

\begin{itemize}
\item {\em Finding the X-factor.} It is often hard to predict what would boost a detection algorithm's performance the most; e.g., better priors or better and/or more (human) labels in learning algorithms, better parameter tuning, creating frameworks that combine multiple algorithms working in parallel or sequentially, choosing the appropriate algorithms for the framework, 
or simply having more data.

\item {\em Evaluation.} Due to the challenges associated with collecting true labels related to cost and annotator noise, ground truth data is often inexistent. As such, various works employ different approaches as was discussed in the Concluding Remarks after Sections \ref{sec:static} and \ref{sec:dynamic}, such as anomaly injections and qualitative analysis.
Thus far, there is no standard for evaluating (graph) anomaly detection methods. 

\item {\em Graph Construction.} Often times the data does not form a network as it is the case in computer networks. Rather, it is up to the practitioners to build a network representation of their data in order to use graph-based techniques. In such cases, it is often hard to anticipate what source of data is best to use in graph construction.

\item {\em Anomaly Detection on Multi-Graphs.} On the contrary to above, it may be the case that there is more than one network available, capturing different aspects of relations (e.g. friendship network and telecommunication network among the same individuals). While possibly beneficial, how to exploit all available networks and fuse clues from all these sources  for anomaly detection remains an open area.

\item {\em Balance between Attribution and `Novelty' Detection.} By anomaly attribution, we essentially want to attribute the detected anomalies to known, human-understandable evidences (e.g., the known frauds, the known rule-based meta detectors, etc). This might contradict to some anomaly detection tasks where the goal is to find `novel' patterns beyond users' current understandings. More research needs to be done in the direction of how to balance between the attribution and the ability of the detection algorithm to find `novelty'.
\item {\em Augmented Graph Anomaly Detection.} When there is an explicit network representation, it may also be possible to introduce/remove {\em latent} edges, for example edges based on (e.g. text, time-series, correlation) similarities or domain knowledge (e.g. known irrelevant types of edges).

\end{itemize}


\begin{acknowledgements}

This material is based upon work supported by  the Army Research Office  (ARO) under Cooperative Agreement Numbers W911NF-14-1-0029 and W911NF-09-2-0053, the Defense Advanced Research Projects Agency (DARPA) under Contract Numbers W911NF-11-C-0088, W911NF-11-C-0200 and W911NF-12-C-0028,
the National Science Foundation (NSF) under Grants No. IIS-1217559 and IIS1017415, by Region II University Transportation Center under the project number 49997-33-25, and the Stony Brook University Office of Vice President for Research.

Any findings and conclusions expressed in this material are those of the author(s) and do not necessarily reflect the position or the policy of the U.S. Government and the other funding parties, and no official endorsement should be inferred.  The U.S. Government is authorized to reproduce and distribute reprints for Government purposes notwithstanding any copyright notation here on.
\end{acknowledgements}

\bibliographystyle{plainnat}
\bibliography{abbreviations,all}

\end{document}